\documentclass[12pt]{iopart}

\usepackage[english]{babel}
\usepackage{multirow}
\usepackage{adjustbox}
\usepackage{tabularx} 
\usepackage{float} 
\usepackage{graphicx}
\usepackage{epstopdf}
\usepackage[dvipsnames]{xcolor} 
\usepackage{dcolumn}
\usepackage{bm}
\usepackage[hidelinks, bookmarks=true]{hyperref}
\usepackage[mathlines]{lineno}
\usepackage{xfrac}
\usepackage{booktabs} 
\usepackage[separate-uncertainty,retain-explicit-plus,per-mode=symbol,binary-units,parse-numbers=false]{siunitx}
\usepackage{enumitem}
\usepackage{csquotes}
\usepackage{stmaryrd}
\usepackage[capitalise,noabbrev]{cleveref} 
\usepackage[mathscr]{euscript}
\usepackage[nice]{nicefrac}
\usepackage{footmisc}
\usepackage[normalem]{ulem}

\usepackage{cite}
\crefformat{equation}{(#2#1#3)}

\newcommand{\dquotes}[1]{``#1''} 

\newcommand\NDBD{$0\nu\beta\beta$}
\newcommand\DBD{$2\nu\beta\beta$}

\newcommand{\addcauthor}[4]{#1$^{{#2},}$\footnotemark[{#3}]}

\begin{document}

\title[]{Neutrinoless Double Beta Decay Sensitivity of the XLZD Rare Event Observatory}

\author{
{\large{The XLZD Collaboration}}  \vspace{1.2mm} \\
J.~Aalbers$^{1}$,
K.~Abe$^{2}$,
M.~Adrover$^{3}$,
S.~Ahmed~Maouloud$^{4}$,
D.~S.~Akerib$^{5,6}$,
A.~K.~Al~Musalhi$^{7}$,
F.~Alder$^{7}$,
L.~Althueser$^{8}$,
D.~W.~P.~Amaral$^{9}$,
C.~S.~Amarasinghe$^{10}$,
A.~Ames$^{5,6}$,
B.~Andrieu$^{4}$,
N.~Angelides$^{11}$,
E.~Angelino$^{12,13}$,
\addtocounter{footnote}{4}
\addcauthor{B.~Antunovic}{14}{0}{Also at University of Banja Luka, 78000 Banja Luka, Bosnia and Herzegovina}, 
E.~Aprile$^{15}$,
\addtocounter{footnote}{-4}
\addcauthor{H.M.~Ara\'{u}jo}{11}{1},
J.~E.~Armstrong$^{16}$,
M.~Arthurs$^{5,6}$,
M.~Babicz$^{3}$,
A.~Baker$^{18,11}$,
M.~Balzer$^{19}$,
J.~Bang$^{20}$,
E.~Barberio$^{21}$,
J.~W.~Bargemann$^{10}$,
E.~Barillier$^{3}$,
A.~Basharina-Freshville$^{7}$,
L.~Baudis$^{3}$,
D.~Bauer$^{11}$,
M.~Bazyk$^{22,21}$,
K.~Beattie$^{23}$,
N.~Beaupere$^{22}$,
N.~F.~Bell$^{21}$,
L.~Bellagamba$^{24}$,
T.~Benson$^{25}$,
A.~Bhatti$^{16}$,
T.~P.~Biesiadzinski$^{5,6}$,
R.~Biondi$^{26}$,
Y.~Biondi$^{27}$,
H.~J.~Birch$^{3}$,
E.~Bishop$^{28}$,
A.~Bismark$^{3}$,
C.~Boehm$^{29}$,
K.~Boese$^{26}$,
A.~Bolotnikov$^{30}$,
P.~Br\'{a}s$^{31}$,
R.~Braun$^{8}$,
A.~Breskin$^{32}$,
C.~A.~J.~Brew$^{33}$,
S.~Brommer$^{34}$,
A.~Brown$^{35,36}$,
G.~Bruni$^{24}$,
R.~Budnik$^{32}$,
S.~Burdin$^{37}$,
C.~Cai$^{38}$,
C.~Capelli$^{3}$,
G.~Carini$^{30}$,
M.~C.~Carmona-Benitez$^{39}$,
M.~Carter$^{37}$,
A.~Chauvin$^{40}$,
A.~Chawla$^{41}$,
H.~Chen$^{23}$,
J.~J.~Cherwinka$^{25}$,
Y.~T.~Chin$^{39}$,
N.~I.~Chott$^{42}$,
A.~P.~Cimental~Chavez$^{3}$,
K.~Clark$^{43}$,
A.~P.~Colijn$^{44}$,
D.~J.~Colling$^{11}$,
J.~Conrad$^{45}$,
M.~V.~Converse$^{46}$,
R.~Coronel$^{5,6}$,
D.~Costanzo$^{36}$,
A.~Cottle$^{7}$,
G.~Cox$^{47}$,
J.~J.~Cuenca-Garc\'ia$^{3}$,
D.~Curran$^{47}$,
D.~Cussans$^{43}$,
\addtocounter{footnote}{5}
\addcauthor{V.~D'Andrea}{13}{6}{Also at INFN-Roma Tre, 00146 Roma, Italy}, \addtocounter{footnote}{1}
L.~C.Daniel~Garcia$^{4}$,
I.~Darlington$^{7}$,
S.~Dave$^{7}$,
A.~David$^{7}$,
G.~J.~Davies$^{11}$,
M.~P.~Decowski$^{44}$,
A.~Deisting$^{48}$,
J.~Delgaudio$^{47}$,
S.~Dey$^{49}$,
C.~Di~Donato$^{50}$,
L.~Di~Felice$^{11}$,
P.~Di~Gangi$^{24}$,
S.~Diglio$^{22}$,
C.~Ding$^{20}$,
J.~E.~Y.~Dobson$^{18}$,
M.~Doerenkamp$^{40}$,
G.~Drexlin$^{34}$,
E.~Druszkiewicz$^{46}$,
C.~L.~Dunbar$^{47}$,
K.~Eitel$^{27}$,
A.~Elykov$^{27}$,
R.~Engel$^{27}$,
S.~R.~Eriksen$^{43}$,
S.~Fayer$^{11}$,
N.~M.~Fearon$^{49}$,
A.~D.~Ferella$^{50,13}$,
C.~Ferrari$^{13}$,
N.~Fieldhouse$^{49}$,
H.~Fischer$^{35}$,
H.~Flaecher$^{43}$,
T.~Flehmke$^{45}$,
M.~Flierman$^{44}$,
E.~D.~Fraser$^{37}$,
T.M.A.~Fruth$^{29}$,
K.~Fujikawa$^{51}$,
W.~Fulgione$^{12,13}$,
C.~Fuselli$^{44}$,
P.~Gaemers$^{44}$,
R.~Gaior$^{4}$,
R.~J.~Gaitskell$^{20}$,
N.~Gallice$^{30}$,
M.~Galloway$^{3}$,
F.~Gao$^{38}$,
N.~Garroum$^{4}$,
A.~Geffre$^{47}$,
J.~Genovesi$^{39}$,
C.~Ghag$^{7}$,
S.~Ghosh$^{5,6,52}$,
R.~Giacomobono$^{53}$,
R.~Gibbons$^{54,23}$,
F.~Girard$^{4}$,
R.~Glade-Beucke$^{35}$,
F.~Gl\"uck$^{27}$,
S.~Gokhale$^{30}$,
L.~Grandi$^{55}$,
J.~Green$^{49}$,
J.~Grigat$^{35}$,
M.~G.~D.~van~der~Grinten$^{33}$,
R.~Gr\"o{\ss}le$^{27}$,
H.~Guan$^{52}$,
M.~Guida$^{26}$,
P.~Gyorgy$^{48}$,
J.~J.~Haiston$^{42}$,
C.~R.~Hall$^{16}$,
T.~Hall$^{37}$,
R.~Hammann$^{26}$,
V.~Hannen$^{8}$,
S.~Hansmann-Menzemer$^{40}$,
N.~Hargittai$^{32}$,
E.~Hartigan-O'Connor$^{20}$,
S.~J.~Haselschwardt$^{56}$,
M.~Hernandez$^{3}$,
S.~A.~Hertel$^{57}$,
A.~Higuera$^{9}$,
C.~Hils$^{48}$,
K.~Hiraoka$^{51}$,
L.~Hoetzsch$^{26}$,
M.~Hoferichter$^{58}$,
G.~J.~Homenides$^{17}$,
N.~F.~Hood$^{59}$,
M.~Horn$^{47}$,
D.~Q.~Huang$^{60}$,
S.~Hughes$^{37}$,
D.~Hunt$^{49}$,
M.~Iacovacci$^{53}$,
Y.~Itow$^{51}$,
E.~Jacquet$^{11}$,
J.~Jakob$^{8}$,
R.~S.~James$^{21}$,
F.~Joerg$^{26,3}$,
S.~Jones$^{36}$,
A.~C.~Kaboth$^{41}$,
F.~Kahlert$^{52}$,
A.~C.~Kamaha$^{60}$,
Y.~Kaminaga$^{2}$,
M.~Kara$^{27}$,
P.~Kavrigin$^{32}$,
S.~Kazama$^{51}$,
M.~Keller$^{40}$,
P.~Kemp-Russell$^{36}$,
D.~Khaitan$^{46}$,
P.~Kharbanda$^{44}$,
B.~Kilminster$^{3}$,
J.~Kim$^{10}$,
R.~Kirk$^{20}$,
M.~Kleifges$^{19}$,
M.~Klute$^{34}$,
M.~Kobayashi$^{51}$,
D.~Kodroff$^{23}$,
D.~Koke$^{8}$,
A.~Kopec$^{61}$,
E.~V.~Korolkova$^{36}$,
H.~Kraus$^{49}$,
S.~Kravitz$^{62}$,
L.~Kreczko$^{43}$,
B.~von~Krosigk$^{63}$,
V.~A.~Kudryavtsev$^{36}$,
\addtocounter{footnote}{-5}
\addcauthor{F.~Kuger}{35}{2}, 
N.~Kurita$^{5}$,
H.~Landsman$^{32}$,
R.~F.~Lang$^{52}$,
C.~Lawes$^{18}$,
J.~Lee$^{64}$,
B.~Lehnert$^{65}$,
D.~S.~Leonard$^{64}$,
K.~T.~Lesko$^{23}$,
L.~Levinson$^{32}$,
A.~Li$^{59}$,
I.~Li$^{9}$,
S.~Li$^{66}$,
S.~Liang$^{9}$,
Z.~Liang$^{67}$,
J.~Lin$^{54,23}$,
Y.~-T.~Lin$^{26}$,
S.~Lindemann$^{35}$,
S.~Linden$^{30}$,
M.~Lindner$^{26}$,
\addtocounter{footnote}{1}
\addcauthor{A.~Lindote}{31}{3}, 
W.~H.~Lippincott$^{10}$,
K.~Liu$^{38}$,
J.~Loizeau$^{22}$,
F.~Lombardi$^{48}$,
\addtocounter{footnote}{4}
\addcauthor{J.~A.~M.~Lopes}{68}{7}{Also at Coimbra Polytechnic - ISEC, 3030-199 Coimbra, Portugal}, \addtocounter{footnote}{1}
M.~I.~Lopes$^{31}$,
W.~Lorenzon$^{56}$,
M.~Loutit$^{43}$,
C.~Lu$^{20}$,
G.~M.~Lucchetti$^{24}$,
T.~Luce$^{35}$,
S.~Luitz$^{5,6}$,
Y.~Ma$^{59}$,
C.~Macolino$^{50,13}$,
J.~Mahlstedt$^{45}$,
B.~Maier$^{34,11}$,
P.~A.~Majewski$^{33}$,
A.~Manalaysay$^{23}$,
A.~Mancuso$^{24}$,
L.~Manenti$^{29}$,
R.~L.~Mannino$^{69}$,
F.~Marignetti$^{53}$,
T.~Marley$^{11}$,
T.~Marrod\'an~Undagoitia$^{26}$,
K.~Martens$^{2}$,
J.~Masbou$^{22}$,
E.~Masson$^{4}$,
S.~Mastroianni$^{53}$,
C.~Maupin$^{47}$,
C.~McCabe$^{18}$,
M.~E.~McCarthy$^{46}$,
D.~N.~McKinsey$^{54,23}$,
J.~B.~Mclaughlin$^{7}$,
A.~Melchiorre$^{50}$,
J.~Men\'endez$^{70}$,
M.~Messina$^{13}$,
E.~H.~Miller$^{5,6}$,
B.~Milosovic$^{14}$,
S.~Milutinovic$^{14}$,
K.~Miuchi$^{71}$,
R.~Miyata$^{51}$,
E.~Mizrachi$^{69,16}$,
A.~Molinario$^{12}$,
C.~M.~B.~Monteiro$^{68}$,
M.~E.~Monzani$^{5,6,72}$,
K.~Mor\aa$^{15}$,
S.~Moriyama$^{2}$,
E.~Morrison$^{42}$,
E.~Morteau$^{22}$,
Y.~Mosbacher$^{32}$,
B.~J.~Mount$^{73}$,
J.~M\"uller$^{35}$,
M.~Murdy$^{57}$,
A.~St.~J.~Murphy$^{28}$,
M.~Murra$^{15}$,
A.~Naylor$^{36}$,
H.~N.~Nelson$^{10}$,
F.~Neves$^{31}$,
J.~L.~Newstead$^{21}$,
A.~Nguyen$^{28}$,
K.~Ni$^{59}$,
C.~O'Hare$^{29}$,
U.~Oberlack$^{48}$,
M.~Obradovic$^{14}$,
\addtocounter{footnote}{-4}
\addcauthor{I.~Olcina}{54,23}{4}{},
K.~C.~Oliver-Mallory$^{11}$,
G.~D.~Orebi~Gann$^{54,23}$,
J.~Orpwood$^{36}$,
S.~Ouahada$^{3}$,
K.~Oyulmaz$^{28}$,
B.~Paetsch$^{32}$,
K.~J.~Palladino$^{49}$,
J.~Palmer$^{41}$,
Y.~Pan$^{4}$,
M.~Pandurovic$^{14}$,
N.~J.~Pannifer$^{43}$,
S.~Paramesvaran$^{43}$,
S.~J.~Patton$^{23}$,
Q.~Pellegrini$^{4}$,
B.~Penning$^{3}$,
G.~Pereira$^{31}$,
R.~Peres$^{3}$,
E.~Perry$^{7}$,
T.~Pershing$^{69}$,
F.~Piastra$^{3}$,
J.~Pienaar$^{32}$,
A.~Piepke$^{17}$,
M.~Pierre$^{44}$,
G.~Plante$^{15}$,
T.~R.~Pollmann$^{44}$,
L.~Principe$^{22,21}$,
J.~Qi$^{59}$,
K.~Qiao$^{44}$,
Y.~Qie$^{46}$,
J.~Qin$^{9}$,
S.~Radeka$^{30}$,
V.~Radeka$^{30}$,
M.~Rajado$^{3}$,
D.~Ram\'irez~Garc\'ia$^{3}$,
A.~Ravindran$^{22,21}$,
A.~Razeto$^{13}$,
J.~Reichenbacher$^{42}$,
C.~A.~Rhyne$^{20}$,
A.~Richards$^{11}$,
G.~R.~C.~Rischbieter$^{56,3}$,
H.~S.~Riyat$^{28}$,
R.~Rosero$^{30}$,
A.~Roy$^{11}$,
T.~Rushton$^{36}$,
D.~Rynders$^{47}$,
R.~Saakyan$^{7}$,
L.~Sanchez$^{9}$,
\addtocounter{footnote}{4}
\addcauthor{P.~Sanchez-Lucas}{3}{8}{Also at University of Grenada},
D.~Santone$^{41}$,
J.~M.~F.~dos~Santos$^{68}$,
G.~Sartorelli$^{24}$,
A.~B.~M.~R.~Sazzad$^{17}$,
A.~Scaffidi$^{74}$,
R.~W.~Schnee$^{42}$,
J.~Schreiner$^{26}$,
P.~Schulte$^{8}$,
H.~Schulze~Ei{\ss}ing$^{8}$,
M.~Schumann$^{35}$,
A.~Schwenck$^{27}$,
A.~Schwenk$^{75,26}$,
L.~Scotto~Lavina$^{4}$,
M.~Selvi$^{24}$,
F.~Semeria$^{24}$,
P.~Shagin$^{48}$,
S.~Sharma$^{40}$,
S.~Shaw$^{28}$,
W.~Shen$^{40}$,
L.~Sherman$^{5,6}$,
S.~Shi$^{56}$,
S.~Y.~Shi$^{15}$,
T.~Shimada$^{51}$,
T.~Shutt$^{5,6}$,
J.~J.~Silk$^{16}$,
C.~Silva$^{27}$,
H.~Simgen$^{26}$,
G.~Sinev$^{42}$,
R.~Singh$^{52}$,
J.~Siniscalco$^{7}$,
M.~Solmaz$^{63,34}$,
V.~N.~Solovov$^{31}$,
Z.~Song$^{67}$,
P.~Sorensen$^{23}$,
J.~Soria$^{54,23}$,
O.~Stanley$^{21,22}$,
M.~Steidl$^{27}$,
T.~Stenhouse$^{7}$,
A.~Stevens$^{35}$,
K.~Stifter$^{5,6}$,
T.~J.~Sumner$^{11}$,
A.~Takeda$^{2}$,
P.-L.~Tan$^{45}$,
D.~J.~Taylor$^{47}$,
W.~C.~Taylor$^{20}$,
D.~Thers$^{22}$,
T.~Th\"ummler$^{27}$,
D.~R.~Tiedt$^{47}$,
F.~T\"onnies$^{35}$,
Z.~Tong$^{11}$,
F.~Toschi$^{27}$,
D.~R.~Tovey$^{36}$,
J.~Tranter$^{36}$,
M.~Trask$^{10}$,
G.~Trinchero$^{12}$,
M.~Tripathi$^{76}$,
D.~R.~Tronstad$^{42}$,
R.~Trotta$^{74,11}$,
C.~D.~Tunnell$^{9}$,
P.~Urquijo$^{21}$,
A.~Us\'{o}n$^{28}$,
M.~Utoyama$^{51}$,
A.~C.~Vaitkus$^{20}$,
O.~Valentino$^{11}$,
K.~Valerius$^{27}$,
S.~Vecchi$^{77}$,
V.~Velan$^{23}$,
S.~Vetter$^{27}$,
L.~de~Viveiros$^{39}$,
G.~Volta$^{26}$,
D.~Vorkapic$^{14}$,
A.~Wang$^{5,6}$,
J.~J.~Wang$^{17}$,
Y.~Wang$^{54,23}$,
D.~Waters$^{7}$,
K.~M.~Weerman$^{44}$,
C.~Weinheimer$^{8}$,
M.~Weiss$^{32}$,
D.~Wenz$^{8}$,
T.~J.~Whitis$^{10}$,
K.~Wild$^{39}$,
M.~Williams$^{54,23}$,
M.~Wilson$^{27}$,
S.~T.~Wilson$^{36}$,
C.~Wittweg$^{3}$,
J.~Wolf$^{34}$,
F.~L.~H.~Wolfs$^{46}$,
S.~Woodford$^{37}$,
D.~Woodward$^{23}$,
M.~Worcester$^{30}$,
C.~J.~Wright$^{43}$,
V.~H.~S.~Wu$^{27}$,
S.~W\"ustling$^{19}$,
M.~Wurm$^{48}$,
Q.~Xia$^{23}$,
Y.~Xing$^{21}$,
D.~Xu$^{15}$,
J.~Xu$^{69}$,
Y.~Xu$^{60}$,
Z.~Xu$^{15}$,
M.~Yamashita$^{2}$,
L.~Yang$^{59}$,
J.~Ye$^{67}$,
M.~Yeh$^{30}$,
B.~Yu$^{30}$,
G.~Zavattini$^{77}$,
W.~Zha$^{39}$,
M.~Zhong$^{59}$
and
K.~Zuber$^{65}$
}

\maketitle

\address{$^{1}$~Nikhef and the University of Groningen, Van Swinderen Institute, 9747AG Groningen, Netherlands}
\address{$^{2}$~Kamioka Observatory, Institute for Cosmic Ray Research, and Kavli Institute for the Physics and Mathematics of the Universe (WPI), University of Tokyo, Higashi-Mozumi, Kamioka, Hida, Gifu 506-1205, Japan}
\address{$^{3}$~Physik-Institut, University of Z\"urich, 8057  Z\"urich, Switzerland}
\address{$^{4}$~LPNHE, Sorbonne Universit\'{e}, CNRS/IN2P3, 75005 Paris, France}
\address{$^{5}$~SLAC National Accelerator Laboratory, Menlo Park, CA 94025-7015, USA}
\address{$^{6}$~Kavli Institute for Particle Astrophysics and Cosmology, Stanford University, Stanford, CA  94305-4085 USA}
\address{$^{7}$~Department of Physics and Astronomy, University College London (UCL), London WC1E 6BT, UK}
\address{$^{8}$~Institute for Nuclear Physics, University of M\"unster, 48149 M\"unster, Germany}
\address{$^{9}$~Department of Physics and Astronomy, Rice University, Houston, TX 77005, USA}
\address{$^{10}$~Department of Physics, University of California, Santa Barbara,  Santa Barbara, CA 93106-9530, USA}
\address{$^{11}$~Department of Physics, Imperial College London, Blackett Laboratory, London SW7 2AZ, UK}
\address{$^{12}$~INAF-Astrophysical Observatory of Torino, Department of Physics, University  of  Torino and  INFN-Torino,  10125  Torino,  Italy}
\address{$^{13}$~INFN-Laboratori Nazionali del Gran Sasso and Gran Sasso Science Institute, 67100 L'Aquila, Italy}
\address{$^{14}$~Vinca Institute of Nuclear Science, University of Belgrade, Mihajla Petrovica Alasa 12-14. Belgrade, Serbia}
\address{$^{15}$~Physics Department, Columbia University, New York, NY 10027, USA}
\address{$^{16}$~Department of Physics, University of Maryland, College Park, MD 20742-4111, USA}
\address{$^{17}$~Department of Physics \& Astronomy, University of Alabama, Tuscaloosa, AL 34587-0324, USA}
\address{$^{18}$~Department of Physics, King's College London, London WC2R 2LS, UK}
\address{$^{19}$~Institute for Data Processing and Electronics, Karlsruhe Institute of Technology, 76021 Karlsruhe, Germany}
\address{$^{20}$~Department of Physics, Brown University, Providence, RI 02912-9037, USA}
\address{$^{21}$~ARC Centre of Excellence for Dark Matter Particle Physics, School of Physics, The University of Melbourne, VIC 3010, Australia}
\address{$^{22}$~SUBATECH, IMT Atlantique, CNRS/IN2P3,  Nantes Universit\'e, Nantes 44307, France}
\address{$^{23}$~Lawrence Berkeley National Laboratory (LBNL), Berkeley, CA 94720-8099, USA}
\address{$^{24}$~Department of Physics and Astronomy, University of Bologna and INFN-Bologna, 40126 Bologna, Italy}
\address{$^{25}$~Physical Sciences Laboratory, University of Wisconsin-Madison, Madison, WI 53589-3034, USA}
\address{$^{26}$~Max-Planck-Institut f\"ur Kernphysik, 69117 Heidelberg, Germany}
\address{$^{27}$~Institute for Astroparticle Physics, Karlsruhe Institute of Technology, 76021 Karlsruhe, Germany}
\address{$^{28}$~SUPA, School of Physics and Astronomy, University of Edinburgh, Edinburgh,  EH9 3FD, UK}
\address{$^{29}$~School of Physics, The University of Sydney, Camperdown, Sydney, NSW 2006, Australia}
\address{$^{30}$~Brookhaven National Laboratory (BNL), Upton, NY 11973-5000, USA}
\address{$^{31}$~Laborat\'orio de Instrumenta\c c\~ao e F\'isica Experimental de Part\'iculas (LIP), University of Coimbra, P-3004 516 Coimbra, Portugal}
\address{$^{32}$~Department of Particle Physics and Astrophysics, Weizmann Institute of Science, Rehovot 7610001, Israel}
\address{$^{33}$~STFC Rutherford Appleton Laboratory (RAL), Didcot, OX11 0QX, UK}
\address{$^{34}$~Institute of Experimental Particle Physics, Karlsruhe Institute of Technology, 76021 Karlsruhe, Germany}
\address{$^{35}$~Physikalisches Institut, Universit\"at Freiburg, 79104 Freiburg, Germany}
\address{$^{36}$~School of Mathematical and Physical Sciences, University of Sheffield, Sheffield S3 7RH, UK}
\address{$^{37}$~Department of Physics, University of Liverpool, Liverpool L69 7ZE, UK}
\address{$^{38}$~Department of Physics \& Center for High Energy Physics, Tsinghua University, Beijing 100084, P.R. China}
\address{$^{39}$~Department of Physics, Pennsylvania State University, University Park, PA 16802-6300, USA}
\address{$^{40}$~Physikalisches Institut, Universit\"at Heidelberg, Heidelberg, Germany}
\address{$^{41}$~Department of Physics, Royal Holloway, University of London, Egham, TW20 0EX, UK}
\address{$^{42}$~South Dakota School of Mines and Technology, Rapid City, SD 57701-3901, USA}
\address{$^{43}$~H.H. Wills Physics Laboratory, University of Bristol, Bristol, BS8 1TL, UK}
\address{$^{44}$~Nikhef and the University of Amsterdam, Science Park, 1098XG Amsterdam, Netherlands}
\address{$^{45}$~Oskar Klein Centre, Department of Physics, Stockholm University, AlbaNova, Stockholm SE-10691, Sweden}
\address{$^{46}$~Department of Physics and Astronomy, University of Rochester, Rochester, NY 14627-0171, USA}
\address{$^{47}$~South Dakota Science and Technology Authority (SDSTA), Sanford Underground Research Facility, Lead, SD 57754-1700, USA}
\address{$^{48}$~Institut f\"ur Physik \& Exzellenzcluster PRISMA$^{+}$, Johannes Gutenberg-Universit\"at Mainz, 55099 Mainz, Germany}
\address{$^{49}$~Department of Physics, University of Oxford, Oxford OX1 3RH, UK}
\address{$^{50}$~Department of Physics and Chemistry, University of L'Aquila, 67100 L'Aquila, Italy}
\address{$^{51}$~Kobayashi-Maskawa Institute for the Origin of Particles and the Universe, and Institute for Space-Earth Environmental Research, Nagoya University, Furo-cho, Chikusa-ku, Nagoya, Aichi 464-8602, Japan}
\address{$^{52}$~Department of Physics and Astronomy, Purdue University, West Lafayette, IN 47907, USA}
\address{$^{53}$~Department of Physics ``Ettore Pancini'', University of Napoli and INFN-Napoli, 80126 Napoli, Italy}
\address{$^{54}$~Department of Physics, University of California, Berkeley,  Berkeley, CA 94720-7300, USA}
\address{$^{55}$~Department of Physics, Enrico Fermi Institute \& Kavli Institute for Cosmological Physics, University of Chicago, Chicago, IL 60637, USA}
\address{$^{56}$~Randall Laboratory of Physics, University of Michigan, Ann Arbor, MI 48109-1040, USA}
\address{$^{57}$~Department of Physics, University of Massachusetts, Amherst, MA 01003-9337, USA}
\address{$^{58}$~Albert Einstein Center for Fundamental Physics, Institute for Theoretical Physics, University of Bern, Sidlerstrasse 5, 3012 Bern, Switzerland}
\address{$^{59}$~Department of Physics, University of California San Diego, La Jolla, CA 92093, USA}
\address{$^{60}$~Department of Physics \& Astronomy, University of Califonia, Los Angeles, Los Angeles, CA 90095-1547}
\address{$^{61}$~Department of Physics \& Astronomy, Bucknell University, Lewisburg, PA, USA}
\address{$^{62}$~Department of Physics, University of Texas at Austin, Austin, TX 78712-1192, USA}
\address{$^{63}$~Kirchhoff-Institut f\"ur Physik, Universit\"at Heidelberg, Heidelberg, Germany}
\address{$^{64}$~IBS Center for Underground Physics (CUP), Yuseong-gu, Daejeon, Korea}
\address{$^{65}$~Institut f\"ur Kern und Teilchenphysik, Technische Universit\"at Dresden, 01069 Dresden, Germany}
\address{$^{66}$~Department of Physics, School of Science, Westlake University, Hangzhou 310030, P.R. China}
\address{$^{67}$~School of Science and Engineering, The Chinese University of Hong Kong (Shenzhen), Shenzhen, Guangdong, 518172, P.R. China}
\address{$^{68}$~LIBPhys, Department of Physics, University of Coimbra, 3004-516 Coimbra, Portugal}
\address{$^{69}$~Lawrence Livermore National Laboratory (LLNL), Livermore, CA 94550-9698, USA}
\address{$^{70}$~Departament de F\'{i}sica Qu\`{a}ntica i Astrof\'{i}sica and Institut de Ci\`{e}ncies del Cosmos, Universitat de Barcelona, 08028 Barcelona, Spain}
\address{$^{71}$~Department of Physics, Kobe University, Kobe, Hyogo 657-8501, Japan}
\address{$^{72}$~Vatican Observatory, Castel Gandolfo, V-00120, Vatican City State}
\address{$^{73}$~School of Natural Sciences, Black Hills State University, Spearfish, SD 57799-0002, USA}
\address{$^{74}$~Theoretical and Scientific Data Science, Scuola Internazionale Superiore di Studi Avanzati (SISSA), 34136 Trieste, Italy}
\address{$^{75}$~Department of Physics, Technische Universit\"at Darmstadt, 64289 Darmstadt, Germany}
\address{$^{76}$~Department of Physics, University of California, Davis, Davis, CA 95616-5270, USA}
\address{$^{77}$~INFN-Ferrara and Dip. di Fisica e Scienze della Terra, Universit\`a di Ferrara, 44122 Ferrara, Italy}

\addtocounter{footnote}{-5}
\footnotetext[3]{\href{mailto:alex@coimbra.lip.pt}{~Corresponding author: alex@coimbra.lip.pt}}
\addtocounter{footnote}{-2}
\footnotetext[1]{\href{mailto:h.araujo@imperial.ac.uk}{~Corresponding author: h.araujo@imperial.ac.uk}}
\addtocounter{footnote}{1}
\footnotetext[2]{\href{mailto:FabianKuger@gmail.com}{~Corresponding author: FabianKuger@gmail.com}}
\addtocounter{footnote}{2}
\footnotetext[4]{\href{mailto:ibles10@berkeley.edu}{~Corresponding author: ibles10@berkeley.edu}}

\addtocounter{footnote}{1}
\footnotetext[1]{~Also at University of Banja Luka, 78000 Banja Luka, Bosnia and Herzegovina}
\addtocounter{footnote}{1}
\footnotetext[6]{~Also at Also at INFN-Roma Tre, 00146 Roma, Italy}
\addtocounter{footnote}{1}
\footnotetext[7]{~Also at Coimbra Polytechnic - ISEC, 3030-199 Coimbra, Portugal}
\addtocounter{footnote}{1}
\footnotetext[8]{~Also at University of Grenada} 

\maketitle

\begin{abstract}
\noindent \noindent The XLZD collaboration is developing a two-phase xenon time projection chamber with an active mass of 60 to \SI{80}{t} capable of probing the remaining WIMP-nucleon interaction parameter space down to the so-called neutrino fog. In this work we show that, based on the performance of currently operating detectors using the same technology and a realistic reduction of radioactivity in detector materials, such an experiment will also be able to competitively search for neutrinoless double beta decay in $^{136}$Xe using a natural-abundance xenon target. XLZD can reach a 3$\sigma$ discovery potential half-life of 5.7$\times$10$^{27}$~yr (and a 90\% CL exclusion of 1.3$\times$10$^{28}$~yr) with 10~years of data taking, corresponding to a Majorana mass range of 7.3--31.3~meV (4.8--20.5~meV). XLZD will thus exclude the inverted neutrino mass ordering parameter space and will start to probe the normal ordering region for most of the nuclear matrix elements commonly considered by the community.
\end{abstract}


\section{\label{sec:intro}Introduction}

The observation of neutrinoless double beta decay (\NDBD) would have far-reaching consequences in Particle Physics and Cosmology. Forbidden by the Standard Model (SM) of Particle Physics, it implies the violation of lepton number as a global conservation law and may establish the Majorana nature of the neutrino~\cite{Majorana_original_paper,PhysRevD.25.2951}. If the decay is mediated by the exchange of a light Majorana neutrino, the corresponding half-life is inversely proportional to the square of the effective Majorana neutrino mass, $\langle m_{\beta\beta}\rangle$,
\begin{equation}
    \label{eq:majorana_mass}
    (T_{1/2}^{0\nu})^{-1} = g_A^4 G^{0\nu}|M^{0\nu}|^2\frac{\langle m_{\beta\beta}\rangle^2}{m_e^2}, 
\end{equation}
where $m_e$ is the electron mass, $g_A$ the axial-vector coupling constant, $G^{0\nu}$ the phase space factor, and $M^{0\nu}$ the nuclear matrix element (NME)~\cite{agostini:2023}.
 
Several experimental techniques have been deployed and more are planned to search for this rare decay in a variety of isotopes~\cite{Dolinski:2019nrj, agostini:2023}, with the current best lower limits (at 90\% CL) on the \NDBD~decay half-life ($T_{1/2}^{0\nu}$) set in the $10^{26}$ yr range, and $\langle m_{\beta\beta}\rangle \leq (28 - 180)$~meV~\cite{kamland-combined-result,gerda-result}. 
The two-phase (liquid/gas) xenon time projection chamber (TPC) is the leading technology in the field of direct search for dark matter in the form of Weakly Interacting Massive Particles (WIMPs)~\cite{xenon-ws,lz-ws,lz_ws_2024}, but it can also be used to search for \NDBD~decay in $^{134}$Xe and $^{136}$Xe~\cite{lz-xe134-sensitivity,lz-xe136-sensitivity,xenon-xe136,darwin}. Efforts are underway to scale up this technology to tens of tonnes of target mass~\cite{Aalbers:2016jon, G3whitepaper:2022}, sufficient to probe the remaining WIMP-nucleon interaction parameter space down to the \dquotes{neutrino fog}~\cite{Billard:2013qya,OHare:2020lva,OHare:2021utq}. The XENON-LUX-ZEPLIN-DARWIN (XLZD) collaboration was formed to consolidate the expertise and resources of the teams using this technology with the goal of building such a detector. The XLZD experiment will lead the dark matter direct-detection field for years to come, but with its large mass, low background and high sensitivity it will also be able to study many other physics channels, such as alternative dark matter candidates and neutrino properties, serving as a rare event observatory~\cite{G3whitepaper:2022,xlzd-designBook}. Amongst these, XLZD will be able to competitively search for \NDBD~in $^{136}$Xe, reaching or even surpassing the sensitivity of current and planned dedicated experiments using this isotope~\cite{kamland-combined-result,nexo-2022,next-sensitivity}.  

In this work, we present the $^{136}$Xe \NDBD~half-life sensitivity projections for XLZD using the expected performance and background rates of the experiment based on those of the current generation of detectors, in particular LUX-ZEPLIN (LZ) and XENONnT. We explore the dependence of this sensitivity on the active xenon mass, muon flux at the host laboratory, and possible $^{136}$Xe enrichment or depletion scenarios. The paper is organized as follows: in \cref{sec:detector} we describe the XLZD experiment; in \cref{sec:backgrounds} we discuss the expected performance of the detector in the most relevant parameters for this search, and the various backgrounds which can impact it, developing two performance scenarios which are used for the sensitivity estimates; \cref{sec:sensitivity} describes the metrics used for calculating the 90\% CL exclusion and 3$\sigma$ discovery half-life sensitivities, with the results for the half-life and Majorana mass reach being presented in \cref{sec:discussion}; \cref{sec:conclusions} offers conclusions from this work.

\vspace{0.5cm}
\section{\label{sec:detector}The XLZD experiment}

The indicative design for the detector at the core of XLZD is a cylindrical, two-phase xenon TPC with an active liquid mass of \SI{60}{t} of natural xenon in a 1:1 height/diameter ratio, corresponding to approximately 3~m in both dimensions~\cite{xlzd-designBook}. A more ambitious scenario is also foreseen in case of a favourable xenon supply market, with an active mass of 80~t in a configuration which maintains the TPC diameter but increases its height to about 4~m. With an abundance of 8.9\% in natural xenon, this corresponds to \SI{5.3}{t} and \SI{7.1}{t} of $^{136}$Xe isotope in each of these mass stages.

Interactions in the active xenon volume lead to the production of a prompt scintillation signal (termed `S1') and a delayed electroluminescence signal (`S2') produced by the ionization electrons drifted upwards to the surface and extracted to the thin gas layer by electric fields defined by the anode, gate, and cathode electrode grids. In the typical configuration of this type of detector, both signals will be detected by two arrays of photosensors, at the top and bottom of the TPC. The standard choice for photosensors in liquid xenon TPCs has been photomultiplier tubes (PMTs) optimised to detect the xenon vacuum ultraviolet (VUV) light, although VUV-optimised Silicon Photomultipliers (SiPMs) or a hybrid configuration of both technologies are also being considered for XLZD. From these signals it is possible to accurately reconstruct the multiplicity, location(s), energy, and recoiling species for the interaction~\cite{xenon_physics,nest}. 
A field cage surrounding the active volume ensures a uniform vertical field and presents a highly reflective polytetrafluoroethylene (PTFE)~\cite{Neves_2017} surface to maximise the collection of light from the S1 scintillation signal (which drives the energy threshold in such detectors). The entire inner TPC will be installed in a double-walled vacuum cryostat for thermal insulation. 

In a design similar to that of LZ~\cite{lz-nim}, the XLZD TPC will be surrounded by two veto systems: the volume of xenon between the TPC field-cage and the inner cryostat wall (dubbed the \dquotes{xenon skin}) will be instrumented with PMTs, and an outer detector (OD) will surround the cryostat with near $4\pi$ coverage. Several design options are being considered for the OD, from the use of liquid scintillator tanks~\cite{lz-nim} to a water-based Cherenkov detector~\cite{XENON:2024wpa}, or a water-based liquid scintillator (WbLS) that would bridge the two media (see Ref.~\cite{xlzd-designBook} for a more detailed discussion on these options). The goal of these two veto systems is to tag neutron and $\gamma$-ray backgrounds of radiogenic or cosmogenic origin with high efficiency~\cite{lz-sims,lz-nim,XENON:2024wpa}.

To shield against cosmic-ray radiation, XLZD must be operated deep underground. Several world-renowned laboratories have indicated interest in hosting the experiment and are being considered for the final installation: Boulby~\cite{boulby}, Kamioka~\cite{kamioka}, LNGS~\cite{lngs}, SNOLAB~\cite{snolab} and SURF~\cite{surf}. They offer varying levels of reduction of the cosmic muon flux, which in the case of \NDBD~search has direct impact in the $^{137}$Xe background, as discussed in \cref{sec:backgrounds}. In \cref{sec:discussion} we discuss the impact on the \NDBD~sensitivity of operating XLZD at each of these laboratories.

The TPC and veto systems will be installed in a large water tank to shield the experiment from environmental $\gamma$-rays and neutrons mainly emitted from the walls of the laboratory due to trace amounts of $^{238}$U and $^{232}$Th in the rock. Simulations have shown that a 4~m water equivalent (w.e.) of shielding on all sides can make these sources of background negligible for both WIMP and \NDBD~searches, resulting in a water tank with 12~m in diameter and height~\cite{jemima_iop}. Many complex ancillary systems are required to support the operation of these detectors, but their description is beyond the scope of this work.  More details on these systems can be found in Ref.~\cite{xlzd-designBook}.

\vspace{0.5cm}
\section{\label{sec:backgrounds}Detector modeling and backgrounds}

At this early (pre-conceptual) stage of the design of XLZD, we assume a performance similar to that of the currently running xenon-based experiments XENONnT and LZ. We describe below the main assumptions regarding the detector parameters that are most relevant for the NDBD search. These assumptions will be revisited as the design evolves.

The energy resolution of the detector at the \NDBD~decay energy ($Q_{\beta\beta}=2457.83\pm0.37$~keV~\cite{xe136-qbb}) is a key parameter, as it drives the leakage of events from nearby background $\gamma$-ray lines into the signal region (see \cref{ssec:gammas}). Detectors based on the two-phase xenon TPC design have demonstrated sub-1\% ($\sigma$) relative energy resolution in this energy region, with XENON1T reporting 0.80\% \cite{Aprile:2020yad} and LZ 0.67\%~\cite{Pereira_2023}. LZ attained this resolution by using only the unsaturated S2 signal collected in the bottom PMT array and applying granular temporal and spatial corrections to the S1 and S2 pulses, obtained from calibration and background signals (alpha decays in the $^{222}$Rn chain, $^{131m}$Xe and $^{83m}$Kr decays, and pulses from single extracted electrons). In this study we consider a slightly more optimistic resolution of 0.65\%, which we believe is within reach for XLZD using a similar approach. We define the Region-of-Interest (ROI) for \NDBD~search as $\pm1\sigma$ around the $Q_{\beta\beta}$ peak, corresponding to a 32~keV-wide interval, (2441.8--2473.8~keV). High-energy $\gamma$-ray lines around $Q_{\beta\beta}$, from $^{214}$Bi and $^{208}$Tl from trace radioactivity in the detector materials, will be used to determine and monitor the energy scale and resolution (see \cref{ssec:gammas}). 

The ability to reject multiple interactions in the same event is also critical, as it allows to significantly reduce the $\gamma$-ray background from external sources with energy in the \NDBD~ROI: MeV-scale $\gamma$-rays are likely to produce one or more Compton scatters before being absorbed, leading to multiple-site (MS) events. By contrast, \NDBD~decays are essentially point-like to a large degree and thus single-site (SS), with the emitted electrons having short tracks, each on the scale of 1~mm in the dense liquid. Critically, the extremely low energy threshold of these detectors makes this rejection very efficient. Bremsstrahlung photons emitted by the fast \NDBD~electrons can nevertheless travel far enough for the decay to be viewed as a multiple interaction, and thus lead to some loss in signal efficiency. In this study we consider SS/MS separation only in the vertical ($z$) axis, and assume that interactions 3~mm or further apart can be rejected. This choice of threshold follows what has been considered in the literature for two-phase xenon TPCs~\cite{lz-xe136-sensitivity, Baudis_2014}, and is further supported by preliminary results from the LZ detector, which show that a separation of \SI{2}{mm} is possible throughout the full detector depth with a applied \SI{193}{V/cm} drift field~\cite{sandros_thesis}. Analysis of Monte Carlo simulations performed for the DARWIN detector~\cite{darwin} using a 3~mm SS/MS discrimination resolution shows that it should be possible to reject 90\% of the $\gamma$-ray background in the ROI while maintaining 85\% efficiency for signal events. Moreover, it also provides some rejection capability (23\%) for electron backgrounds in the ROI (from $\beta$ decays and solar neutrino interactions, see Sections~\ref{ssec:irreducible} -- \ref{ssec:cosmogenic}), leveraging the emission of bremsstrahlung photons by these electrons. Additional capability to identify multiple interactions in the horizontal (xy) plane would contribute to further improvements in the sensitivity, but we do not explore it in this work.

The additional veto systems surrounding the detector help to further extend the capability of detecting multiple interaction events. Coincident energy depositions in the skin or the OD may be caused directly by the particle that interacts in the TPC (on its way in or out) or by other particles emitted simultaneously. In this study we assume that events in which $>$100~keV is deposited in either of the veto systems in prompt coincidence with an interaction in the TPC can be excluded.

\subsection{\label{ssec:gammas}External $\gamma$-rays}

The most important backgrounds in the search for \NDBD~in $^{136}$Xe arise from high-energy $\gamma$-rays emitted in the decay of $^{214}$Bi (\SI{2448}{keV}) and $^{208}$Tl (\SI{2615}{keV}), which are part of the decay chains of $^{238}$U and $^{232}$Th, respectively. These long-lived isotopes are present in trace amounts in the materials used to build the detector and in the surrounding environment -- most notably in the laboratory rock. We assume that the water tank used for XLZD will be large enough to efficiently shield it against high-energy $\gamma$-rays from the rock, rendering this particular source of background negligible.

The 2448~keV $^{214}$Bi line is the most problematic, as it is less than \SI{10}{keV} away from $Q_{\beta\beta}$.
At \SI{2615}{keV} the $^{208}$Tl SS peak is some 9~$\sigma$ away from the ROI, only adding a small contribution from the continuum created by the $\gamma$-rays that lose a fraction of their energy by Compton scattering before entering the TPC. A relative contribution of 65\% and 35\% from these two lines in the \NDBD~ROI has been estimated for DARWIN~\cite{darwin}. 

\begin{figure}[tbp]
    \begin{indented}
        \item[] \includegraphics[width=0.75\columnwidth]{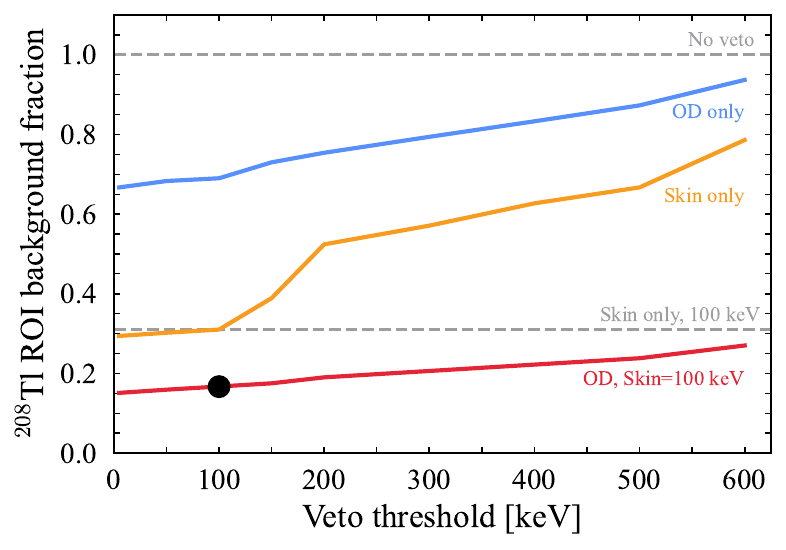}
    \end{indented}
    \caption{\small Simulated fraction of $^{208}$Tl-originated interactions in the central region of the LZ detector that pass the SS/MS and veto cuts and fall inside the \NDBD~ROI, as a function of the energy threshold in each of the veto systems, and of the threshold in the OD for a fixed 100~keV skin threshold. The skin alone can reject close to 70\% of this background with a 100~keV threshold; adding the OD veto with the same threshold further increases this rejection to 83.3\%. The black circle shows the scenario assumed in this study.}
    \label{fig:veto-effect}
\end{figure}

The use of a xenon skin and an OD veto surrounding the TPC can significantly reduce the $^{208}$Tl contribution, by detecting $\gamma$-rays emitted in coincidence with the 2615~keV line (particularly the \SI{583}{keV} $\gamma$-ray with 85\% relative intensity) and/or low energy Compton scatters produced by the 2615~keV $\gamma$-rays themselves in those veto systems before entering the active region of the TPC. In the case of LZ, simulations show that the 
$^{208}$Tl line contributes only on the order of 10\% to the $\gamma$-ray background from detector materials in the \NDBD~ROI for an energy threshold of 100~keV in both the skin and the OD~\cite{lz-xe136-sensitivity}. \cref{fig:veto-effect} shows the effect of the thresholds of both veto detectors on the population of $^{208}$Tl SS events in the  central volume of LZ and for the \NDBD\ ROI defined in this work, obtained from the full background model of the LZ experiment~\cite{lz-sims,lz-backgrounds}. The skin is clearly more powerful for rejecting these events, with the simultaneous use of the OD further reducing this population to approximately 17\% of its initial rate. Consequently, we assume that the (already subdominant) $^{208}$Tl contribution can be efficiently mitigated by the veto systems and focus our attention on the $^{214}$Bi line. 

\begin{figure}[tbp]
    \begin{indented}
        \item[] \includegraphics[width=0.75\columnwidth]{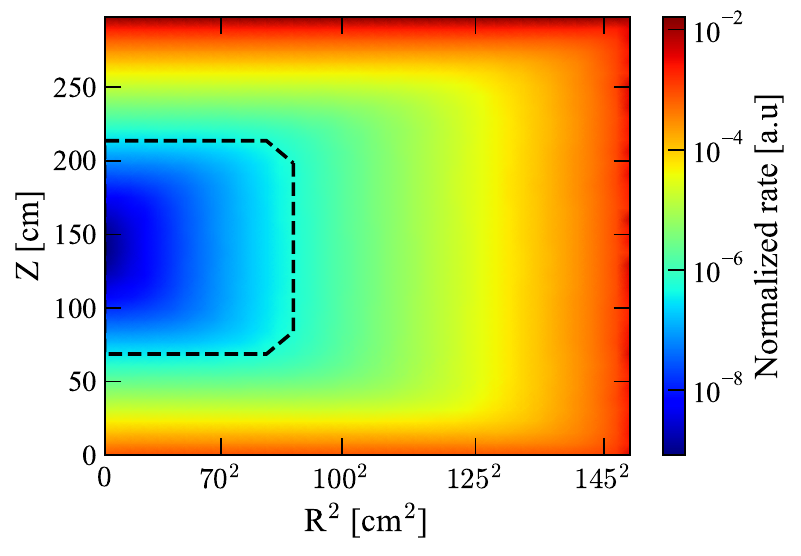}
    \end{indented}
    \caption{\small Example distribution of single scatter events in the \NDBD~ROI caused by interactions of 2448~keV $\gamma$-rays emitted by the decay of $^{214}$Bi in detector materials for the XLZD baseline design of \SI{60}{t}. The effect of xenon self-shielding is clearly visible, with the rate decreasing several orders of magnitude in the central region compared to the surfaces. A higher rate at the top compared to the bottom reflects the additional shielding provided by the xenon in the reverse field region between the cathode and the lower array of photosensors.
    Also shown (black dashed contour) are the limits of the \SI{11}{t} optimal fiducial volume for the 3$\sigma$-significance discovery potential in the optimistic scenario (see discussion in \cref{sec:sensitivity}).
    }
    \label{fig:zr-gammas}
\end{figure}

Large detectors such as XLZD greatly benefit from xenon self-shielding. At the $^{214}$Bi line energy the photoelectric mean interaction length is around 4~m, which is comparable to the dimensions of the TPC. However, $\gamma$-rays must survive many Compton interaction lengths ($\sim$10~cm) without scattering in order to create single-site photoabsorption interactions in the central region. Since such Compton scatters can be detected down to keV energies, this background affects mostly the outer regions of the TPC, creating a highly non-uniform distribution as a result. Simulating this background in the central region of these large volumes with detailed Monte Carlo simulations is extremely challenging computationally. Moreover, it would be impractical to generate detailed simulations of the $\gamma$-ray radiation for all the geometries and scenarios discussed in this work at this early stage of the XLZD project. Instead, we use a semi-empirical model of the attenuation of $^{214}$Bi $\gamma$-rays, emitted from sources uniformly distributed on the surfaces of the TPC, to estimate the relative spatial distributions of this background in the two XLZD mass stages, which are then normalised using the LZ background model~\cite{Robs_thesis}. This model allows for a fast estimation of the $\gamma$-ray background under different scenarios and has been validated against the $\gamma$-ray background profiles of both LZ and DARWIN~\cite{Robs_thesis}. An example of the spatial distribution of SS events in the \NDBD~ROI assuming the \SI{60}{t} detector design as predicted by the semi-empirical model is shown in \cref{fig:zr-gammas}.

To determine the $\gamma$-ray rate normalisation, we assume that the geometrical configuration and construction materials will be similar to those used in LZ. The normalisation factor is calculated by multiplying the simulated rate from SS $^{214}$Bi $\gamma$-rays in LZ by the ratio of the internal surface areas of XLZD and LZ. The second column in \cref{table:backgrounds} shows the detector components with the highest contributions to the LZ \NDBD~background~\cite{lz-xe136-sensitivity}. 

Thorough material selection and cleanliness protocols have always been a concern for direct dark matter search experiments (see, e.g., Refs.~\cite{lz_assays, xen1t_assays, xen_nt_assays}), but these detectors were optimised for the low-energy region, where the xenon self-shielding effect is even more powerful, making the external $\gamma$-ray background subdominant. The material radioactivity requirements for a \NDBD~decay search must be even more stringent.
 
Analyzing the case of LZ, the $\gamma$-ray background in the \NDBD~region could realistically be reduced through a combination of stricter selection of material batches and alternative design choices: for instance, using purpose-built field-shaping resistors (similarly to those described in Ref.~\cite{exo200_assays}); using even lower radioactivity capacitors in the PMT bases, such as those identified in Ref.~\cite{xen1t_assays}; ensuring PMT cables are assembled from low background copper batches under strict cleanliness protocols; avoiding the reuse of higher radioactivity PMTs from LUX and using instead only the cleaner Hamamatsu R8520 1" PMTs in the skin region~\cite{lz_assays}; avoiding dead regions in the OD very close to the cryostat. The contributions from most subsystems could also be significantly reduced if all the parts were fabricated from the cleanest batches of PTFE, stainless steel, copper, Kovar, and Kapton identified during the LZ and XENON assay campaigns~\cite{lz_assays, xen1t_assays}. 

For the particular case of the dominant contribution from the TPC PMTs, Hamamatsu has been working on the development of a lower radioactivity variant of the same model. Prototypes of these new lower radioactivity PMTs have recently been assayed by the XLZD groups, showing a reduction of 2/3 in $^{226}$Ra compared to the original R11410-22 model, and even lower radioactivity PMTs are now becoming available~\cite{pandaX-newPMTs}. The use of lower radioactivity devices such as SiPMs in the top array (alone or in an hybrid configuration including PMTs in the periphery) could provide an additional reduction. 

The third column of \cref{table:backgrounds} shows the impact these changes would have in the projected LZ $^{214}$Bi background. Smaller contributors, grouped under ``Other components'', include the OD acrylic tanks and scintillator, cryostat seals and super-insulation, grid holders and the inner TPC PTFE. Of these we highlight that the mass of stainless steel in the grid rings is minimised by using woven grids, and that clean cryostat super-insulating materials have been identified by other experiments (e.g.~GERDA). Overall, we consider that a reduction of approximately 75\% of the $^{214}$Bi $\gamma$-ray background can be achieved with confidence based entirely on public assay data from the LZ and XENON campaigns and published literature. Moreover, we also assume a similar reduction in the subdominant $^{208}$Tl contribution.

\begin{table}[tbp]
    \caption{Simulated background of single scatter events from $^{214}$Bi 2448~keV $\gamma$-rays originating from the decay of $^{238}$U in detector materials in the LZ experiment. Listed are the highest contributors to the \NDBD~ search, with the corresponding expected number of events in the $\pm$1$\sigma$ ROI (assuming 1.0\% energy resolution) and in the \SI{967}{kg} inner fiducial volume of LZ, for a 1000~live days run~\cite{lz-xe136-sensitivity}. Smaller contributors grouped under ``Other components'' are listed in the text. The third column shows the expected number of events after application of the mitigation strategies described in the text, resulting in an overall radioactivity reduction of approximately 75\%. The fourth column shows the projected events in the \SI{8.2}{t} fiducial volume of a \SI{60}{t} XLZD (see \cref{sec:sensitivity}) in 10~years under this scenario.}
    \label{table:backgrounds} 
    \begin{indented}
        \item[] \begin{tabular}{l c c c}
         \hline\hline
          \rule{0pt}{2.5ex} & \multicolumn{3}{c}{$^{214}$Bi events} \\
          & \multicolumn{2}{c}{LZ} & XLZD \\
          & \multicolumn{2}{c}{(\SI{967}{kg} $\times$ \SI{1000}{d})} & (\SI{8.2}{t} $\times$ \SI{10}{yr})\\
         \rule{0pt}{2.5ex}Component & Nominal & Reduced & Projected  \\ [0.5ex] 
         \hline\hline
         \rule{0pt}{2.5ex}TPC PMTs & 2.95 & 0.98 & 0.14 \\ 
         PMT structures & 2.75 & 0.54 & 0.08 \\
         Field-cage resistors & 2.46 & 0 & 0 \\
         Internal sensors & 1.81 & 0.22 & 0.03 \\
         PMT bases & 1.52 & 0.39 & 0.05 \\
         Cryostat & 1.26 & 0.82 & 0.12 \\
         PMT cables & 1.01 & 0.16 & 0.02 \\
         Field-cage rings & 0.97 & 0.40 & 0.06 \\
         OD tank supports & 0.73 & 0 & 0 \\
         OD foam & 0.71 & 0 & 0 \\
         Skin PMTs & 0.69 & 0.06 & 0.01 \\
         Other skin parts & 0.68 & 0.05 & 0.01 \\
         Other components & 3.56 & 1.42 & 0.20 \\
         \hline
         \rule{0pt}{2.5ex}Total & 21.10 & 5.05 & 0.71 \\
         \hline\hline
    \end{tabular}
    \end{indented}
\end{table}

For the projections presented in this work we use the realistic 75\% reduction for the normalization of the $\gamma$-ray model in our nominal scenario. We also consider a more optimistic scenario with a 90\% reduction of the materials background compared to LZ, which we believe would be possible with a lengthier material selection campaign and more targeted design approaches.

\subsection{\label{ssec:irreducible}Irreducible backgrounds}

Neutrinos interact with electrons in the xenon target via charged and neutral currents, producing an irreducible background~\cite{Billard:2013qya,annrev-astro}. The only neutrinos with sufficient energy to produce interactions near $Q_{\beta\beta}$ and a high enough flux to constitute a relevant background for \NDBD~decay are those originating in $^{8}$B decay in the solar proton-proton (pp) cycle~\cite{annrev-astro}, which have a \SI{16}{MeV} end-point. We consider an overall flux of $^{8}$B neutrinos of $5.25 \times 10^{6}$~/(cm$^{2}\cdot$~s)~\cite{sno_b8}, as recommended in Ref.~\cite{Baxter:2021pqo}, and an electron neutrino survival probability of $P_{ee} = 0.543$, resulting in a \NDBD~ROI rate of $1.76\times10^{-4}$~evt/(t$\cdot$yr$\cdot$keV) before SS/MS discrimination, which is lowered by the \SI{3}{mm} MS rejection cut as described in \cref{sec:backgrounds}.

Another source of irreducible background comes from the standard $^{136}$Xe two-neutrino double-beta decay (\DBD) events, with energies near the $Q_{\beta\beta}$ peak. Some of these events will leak into the \NDBD~ROI due to the finite energy resolution of the detector. In XLZD this background is expected to be subdominant given the excellent energy resolution and the narrow width of the \NDBD~ROI. For a \DBD~half-life of $2.17\times10^{21}$ yr~\cite{exo200-xe136-dbd} and the spectral shape provided in Ref.~\cite{KI_phase_space}, the estimated rate in the \NDBD~ROI is $5.04\times10^{-6}$~evt/(t$\cdot$yr$\cdot$keV).

\subsection{\label{ssec:radon}Radon}

Radon-222 is produced in the decay chain of $^{238}$U and these atoms emanate from detector and gas-system surfaces and mix with the xenon target. Meticulous material selection is crucial to minimise radon emanation, along with active radon reduction techniques during detector operation. XENONnT has achieved a $^{222}$Rn background level of $<$\SI{1}{\micro Bq/kg}, in part by using a high-flow cryogenic distillation column for radon removal capable of turning over the full xenon payload on a time scale comparable to the $^{222}$Rn half-life~\cite{xenon_rnremoval}. A larger system will be developed for XLZD which, together with cryogenic material screening and additional surface treatments to provide radon barriers (e.g.~copper electroplating) already being developed will allow a decrease of the $^{222}$Rn activity by at least one order of magnitude, to \SI{0.1}{\micro Bq/kg}~\cite{xlzd-designBook}.

One of the decay products of $^{222}$Rn is $^{214}$Bi, which decays via beta emission: 19.7\% of these decays proceed to the ground state of $^{214}$Po with a beta end-point energy of \SI{3270}{keV}, and have no accompanying de-excitation $\gamma$-rays (a so-called \dquotes{naked} beta)~\cite{TabRad_v4}. This $^{214}$Bi decay channel produces single-site events with an energy spectrum that overlaps with the \NDBD~ROI. The daughter isotope, $^{214}$Po, has a relatively short half-life of \SI{162.3}{\micro s} before it undergoes alpha emission, producing a clear and distinctive signal in the detector. Observation of the delayed coincidence between the $\beta$ and $\alpha$ interactions (known as \dquotes{BiPo tagging}) will be highly efficient in the long event waveforms of XLZD: a minimum of \SI{2}{ms}, corresponding to a \SI{290}{V/cm} drift field in the \SI{60}{t} (\SI{3}{m} height) configuration~\cite{xlzd-designBook}, encompassing 99.98\% of the BiPo decays. A lower drift field or a taller (\SI{80}{t}) detector will lead to an even higher tagging efficiency. Other $^{214}$Bi decay channels include the emission of $\gamma$-rays from the nuclear de-excitation of $^{214}$Po in coincidence with the beta. In the xenon bulk these decays lead to multi-site events which can be excluded with high efficiency. Conservatively assuming a BiPo tagging efficiency of 99.95\% and a SS/MS vertical discrimination of 3~mm results in a total ROI rate of $9.5\times10^{-5}$~~evt/(t$\cdot$yr$\cdot$keV) from all $^{214}$Bi decays in the bulk, making it subdominant to the irreducible $^8$B background.

Decays of $^{214}$Bi in the xenon outside the active region of the TPC can also contribute to the background if they occur in a non-instrumented detector region and the $\beta$ is accompanied by a coincident \SI{2448}{keV} $\gamma$-ray (1.6\% probability) that reaches the inner TPC. The instrumentation of the lateral and dome regions of the skin in XLZD is expected to minimise this background source, but some small contribution is still expected from the fraction of these decays in which the $\beta$ is below the skin threshold (e.g. 22\% below \SI{100}{keV}) or occur in small volume pockets with poor light collection. Nevertheless, these $\gamma$-rays are external to the TPC and their rate in the central region will be suppressed by several orders of magnitude by the xenon self-shielding effect, as discussed above. 

Another potential source of background comes from positively charged $^{222}$Rn daughters in the liquid bulk that can drift from their production sites and attach to the walls or the cathode grid. Evidence for this effect has been observed in EXO-200, with the $^{214}$Bi activity in the bulk being 88.4\% lower than the $^{222}$Rn activity~\cite{exo200_ions}, and to a smaller extent (44.6\%) in LZ~\cite{lz-backgrounds}. In  a significant fraction (25\%) of the $^{214}$Bi decays in the cathode or the walls both the initial beta and the subsequent alpha from $^{214}$Po are absorbed in the solid surfaces and can lead to single-site \SI{2448}{keV} $\gamma$-ray events in the active region. Two-phase xenon TPCs have low thresholds, of the order of keV, and thus XLZD will be able to reject a significant fraction of these events by observing the $^{210}$Pb nucleus recoiling into the liquid in those cases where the $\alpha$ particle enters the solid material.

These latter two $^{222}$Rn induced background topologies will be studied in more detail as the detector design and simulation evolve.

\subsection{\label{ssec:cosmogenic}Cosmogenic backgrounds}

Despite the protection offered by the rock overburden, muons can reach the detector underground -- with a flux that depends on the specific host laboratory, cf.~\cref{table:cosmogenics} -- and produce backgrounds for \NDBD~search. They produce energetic neutrons directly by muon spallation or in spallation-induced electromagnetic or hadronic cascades in the detector or surrounding materials, that can be captured by $^{136}$Xe leading to the production of $^{137}$Xe. This isotope undergoes beta decay with a half-life of 3.82~minutes, with the energy of the electron extending up to \SI{4173}{keV}~\cite{exo200_cosmogenics}. In 67\% of $^{137}$Xe decays there is no accompanying nuclear de-excitation $\gamma$-ray, and a small fraction of these fall in the narrow ROI for \NDBD~search (1.1\% assuming a 0.65\% energy resolution).

Muon-induced $^{137}$Xe production is subdominant to other mechanisms such as capture of natural radioactivity neutrons on the xenon circulating in the purification and radon removal systems outside of the water tank. Here we assume that this component can be made negligible through careful local shielding (e.g.~using high-density polyethylene or dedicated water tanks) and/or installation of xenon buffer volumes (`decay tanks') inside the main water tank, allowing the $^{137}$Xe activity to subside before entering the TPC.

\begin{table}[tbp]
    \caption{\small Underground laboratory depth, muon flux, and projected rates of $^{137}$Xe production~\cite{darwin_cosmogenic} and of subsequent decays leading to SS-like events in the \NDBD~ROI (considering the nominal scenario, see \cref{ssec:performance}) for the different sites being considered for installing XLZD (listed in order of increasing SS ROI rate). Production rates for Boulby and Kamioka are scaled from LNGS using the respective muon flux ratios. The water equivalent depth and muon flux for Boulby refer to a new proposed laboratory at \SI{1300}{m} and are interpolated from measurements at the existing \SI{1100}{m} laboratory and the projections for a \SI{1400}{m} laboratory studied in Ref.~\cite{boulby-study}.}
    \label{table:cosmogenics} 
    \begin{indented}
        \item[]\begin{tabular}{l c c c c c r}
        \hline\hline
        \rule{0pt}{2.5ex} & \multicolumn{2}{c}{Depth} & $\mu$ flux & $^{137}$Xe rate & SS ROI rate\\
        Site & [m] & [m w.e.] & [/(m$^{2}\cdot$d)] & [/(t$\cdot$yr)] & [evt/(t$\cdot$yr$\cdot$keV)]\\ [0.5ex] 
        \hline\hline
        \rule{0pt}{2.5ex}SNOLAB & 2070 & 5890 & $<$0.3 & 0.007 & 1.29$\times10^{-6}$ \\ 
        SURF & 1490 & 4300 & 4.6 & 0.142 & 2.72$\times10^{-5}$\\
        Boulby & 1300 & 3330 & 14.6 & 0.404 & 7.73$\times10^{-5}$ \\
        LNGS & 1400 & 3800 & 29.7 & 0.822 & 1.57$\times10^{-4}$ \\
        Kamioka & 1000 & 2700 & 128 & 3.54 & 6.78$\times10^{-4}$  \\
        \hline\hline
    \end{tabular}
    \end{indented}
\end{table}

EXO-200, installed at WIPP, measured a $^{137}$Xe production rate of 2.5~atoms/year per kg of $^{136}$Xe with an 80.6\% $^{136}$Xe-enriched target \cite{exo200_cosmogenics}. XLZD will use natural abundance xenon, which has a $^{136}$Xe abundance of only 8.9\%; moreover, $^{136}$Xe has the lowest neutron capture cross section amongst all naturally occurring isotopes \cite{darwin_cosmogenic}, which results in a much smaller $^{137}$Xe production as the other isotopes effectively shield the $^{136}$Xe. In this study we use the recently estimated production rates for DARWIN at LNGS, SURF and SNOLAB~\cite{darwin_cosmogenic}, while the rates for Kamioka and a proposed new facility at \SI{1300}{m} depth at Boulby are estimated by scaling the LNGS projections by the relative muon fluxes. These production rates are summarised in \cref{table:cosmogenics}, along with the rates of SS-like events in the \NDBD~ROI estimated with a simulation of $^{137}$Xe decays. We note that the new $^{137}$Xe production rate estimate for DARWIN at LNGS, and used here, is a factor of approximately 8 smaller compared to that used in the DARWIN \NDBD~sensitivity study~\cite{darwin,G3whitepaper:2022}, as noted in Refs.~\cite{darwin_cosmogenic,darwin-erratum}, with direct impact on the sensitivity.

The rate of this background could be further reduced by vetoing events for a given number of $^{137}$Xe half-lives following the passage of a muon, or using a delayed coincidence between the muon and the $^{136}$Xe$(n,\gamma)^{137}$Xe process (which produces a $\gamma$-ray cascade adding up to \SI{4025}{keV}~\cite{xe136_capture}) to minimise the false-vetoing rate and consequently the detector dead time in laboratories with higher muon flux. Ref.~\cite{darwin_cosmogenic} reports that 95\% of the $^{136}$Xe captures are with neutrons produced by muons crossing the TPC itself, with the neutron capture occurring shortly after the passage of the muon (up to microseconds). The deposition of hundreds of MeV by a high-energy muon in an ultra-sensitive detector such as a LXe-TPC, optimised for keV-scale interactions, is an extremely disruptive event which can affect the optical readout of the detector for an extended period lasting up to seconds. This implies that the use of a delayed coincidence will not be efficient for most of the $^{137}$Xe production. On the other hand, vetoing the detector after each muon has a large impact on the experiment live-time for the shallower laboratories: e.g., vetoing for one $^{137}$Xe half-life (\SI{3.82}{min}) at Boulby would result in a 76\% live-time fraction. Given that $^{137}$Xe is not dominant except in Kamioka, we conservatively assume that no veto will be applied to reduce this background.

Cosmic-ray muons reaching the underground laboratory also lead to the production of lighter radioisotopes by nuclear spallation, i.e.~the fragmentation of xenon nuclei to create radioactive nuclei. Some of these radioisotopes decay with a total energy in the \NDBD~decay ROI, and may have long half-lives of several hours or even days. KamLAND-Zen~\cite{kamland-cosmogenics} reported a xenon spallation rate of isotopes decaying in their wider [2350--2700]~keV ROI of (3.5±0.6)$\times$10$^{-3}$~/(t$\cdot$d) at the Kamioka laboratory (which has a muon flux of 128~/(m$^2\cdot$~d), the highest amongst the candidate laboratories for hosting XLZD). Most of the isotopes observed by KamLAND decay with the emission of a positron and additional $\gamma$-rays, leading to multi-interaction events which will be easily tagged in XLZD and its vetoes. Those isotopes that undergo $\beta^-$ decay (around 15\%) could in principle be more challenging, but the emitted electron is always accompanied by one or more $\gamma$-rays, which will also result in a very high tagging efficiency in XLZD. Given the much narrower (one-tenth) \NDBD~decay ROI of XLZD and its excellent capability to identify multi-site events with high efficiency, we expect this background to be negligible. Installation of XLZD in a laboratory other than Kamioka would result in a further reduction of at least a factor of 4. KamLAND-Zen uses xenon enriched in $^{136}$Xe (91\%) and $^{134}$Xe (9\%), but recent simulations for DARWIN show that the production rates for the isotopes observed in KamLAND-Zen are similar in natural xenon~\cite{darwin_cosmogenic}. Nevertheless, spallation in natural xenon will result in a wider set of radioisotopes; further simulations are underway to more accurately estimate this background.

\subsection{\label{ssec:performance}Performance scenarios}

We consider two scenarios with different detector performance and background considerations for the sensitivity projections presented in this work, which are summarised in \cref{table:scenarios}. The nominal scenario is based on the performance already achieved by currently running detectors, installation at LNGS and a realistic reduction of the $\gamma$-ray background from detector materials (as discussed in \cref{ssec:gammas}). We also consider a more optimistic scenario, with slight improvements in detector performance, installation at SURF, and a more ambitious reduction of the external $\gamma$-ray background. We believe such a scenario is achievable with a very thorough material screening campaign, design and engineering innovations, and further improvements to photosensor radioactivity -- aided by new analysis techniques such as the use of track topology information (e.g. S2 pulse shape deconvolution). The two active mass configurations (60~t and 80~t) being considered by XLZD are studied under both of these scenarios, resulting in a total of four configurations.

\begin{table}[tbp]
    \caption{Summary of the background assumptions and detector performance parameters used in the two scenarios considered for the sensitivity projections in this study. Irreducible backgrounds from $^8$B neutrinos and $^{136}$Xe \DBD~are constant.}
    \label{table:scenarios} 
    \begin{indented}
        \item[]\begin{tabular}{l c c}
        \hline\hline
        \rule{0pt}{2.5ex} & \multicolumn{2}{c}{Scenario}\\
        Parameter & Nominal & Optimistic \\ [0.5ex] 
        \hline\hline
        \rule{0pt}{2.5ex}$^{222}$Rn concentration [$\mu$Bq/kg] & \multicolumn{2}{c}{0.1} \\
        BiPo tagging efficiency [\%] & 99.95 & 99.99 \\
        External $\gamma$-ray [\% LZ] & 25 & 10 \\
        Installation site & LNGS & SURF \\
        Energy resolution [\%] & 0.65 & 0.60 \\ 
        SS/MS vert. separation [mm] & 3 & 2 \\
        \hline\hline
    \end{tabular}
    \end{indented}   
\end{table}

\begin{figure}[tbp]
    \begin{indented}
        \item[] \includegraphics[width=0.75\columnwidth]{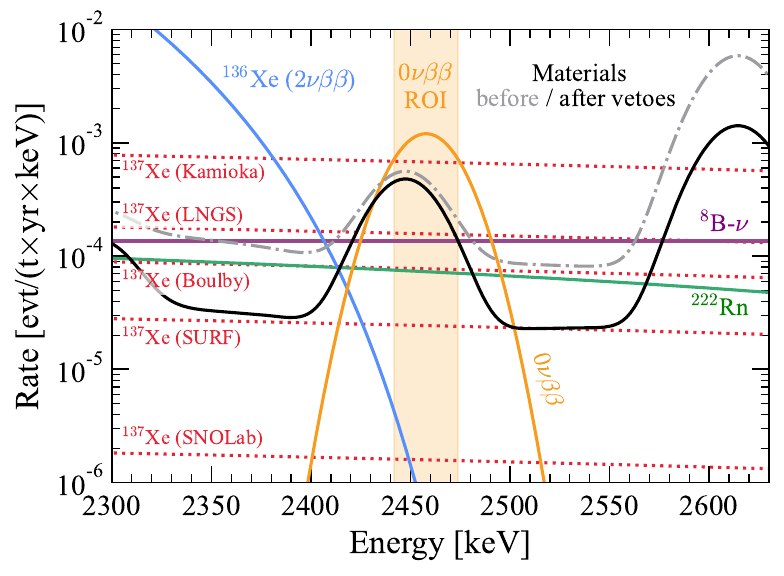}
    \end{indented}
    \caption{Energy spectra of SS events from signal and relevant backgrounds for \NDBD~decay in the inner region of XLZD considering the nominal scenario (see \cref{table:scenarios}). The black (gray) line shows the external $\gamma$-ray background from detector materials after (before) excluding events with coincident signals in the vetoes; their contribution in the \NDBD~ROI (vertical yellow band, defined as $Q_{\beta\beta}\pm1\sigma$ with 0.65\% energy resolution) is clearly dominated by the \SI{2448}{keV} $\gamma$-ray from the decay of $^{214}$Bi in the $^{238}$U chain. The irreducible background from $^8$B solar neutrinos (shown in purple) dominates the internal backgrounds relatively to $^{137}$Xe at the various XLZD candidate host sites (shown by the red dotted lines) except at Kamioka and LNGS (where it is at the same level). $^{222}$Rn (green line) is subdominant even in the nominal scenario, while the contamination from \DBD~decay (light blue) in the ROI is negligible given the excellent energy resolution achieved by LXe-TPCs. Also shown is a hypothetical \NDBD~signal with a half-life of 5$\times$10$^{27}$~yr (orange line).
    }
    \label{fig:bg-model}
\end{figure}

\cref{fig:bg-model} shows the background energy spectra of SS events in the inner region of a \SI{60}{t} XLZD, considering the nominal scenario. $^8$B solar neutrinos dominate the internal backgrounds for all possible installation sites except Kamioka (at LNGS the $^{137}$Xe contribution is at the same level of $^8$B); the effect of $^{222}$Rn is negligible even with the conservative 99.95\% BiPo efficiency considered in this scenario; the power of the excellent energy resolution in minimising the contamination from \DBD~in the ROI is evident. Moreover, this figure clearly shows the impact that the use of external vetoes has on the $^{208}$Tl Compton plateau in the \NDBD~ROI. Vetoed $^{208}$Tl events can nevertheless be used to determine and monitor the energy resolution of the detector, as is done in LZ~\cite{Pereira_2023}, with close to 10$^6$ SS events expected in the $^{208}$Tl peak in the active volume of the \SI{60}{t} stage of XLZD over 10 years of data taking in the nominal scenario ($2\times10^3$ in the inner \SI{30}{t}).

\section{\label{sec:sensitivity}Sensitivity calculation}

We calculate the $^{136}$Xe \NDBD~decay half-life ($T_{1/2}^{0\nu}$) sensitivity using two common metrics: exclusion at 90\% confidence level (CL) and 3$\sigma$-significance discovery potential (defined as the minimum $T_{1/2}^{0\nu}$ required to exclude the null hypothesis with a significance of 99.7\% CL).

Following Ref.~\cite{matteo_fom} we apply a Figure-of-Merit estimator as a straightforward and easily comparable metric for sensitivity calculations. This approach uses a heuristic counting experiment model to determine the signal expectation $S(B)$ required to pass the statistical test in the given metric, assuming a background count $B$. This is converted to the half-life sensitivity using:
\begin{equation}
    \label{eq:figure-of-merit}    
    T_{1/2}^{0\nu} = \ln 2 \frac{N_A \mathscr{E}}{M_{\rm Xe} S(B)}\ ,
\end{equation}
where $\mathscr{E}$ is the sensitive exposure of $^{136}$Xe (i.e. the product of the $^{136}$Xe mass in the fiducial volume, the measurement time, and the signal detection efficiency), $N_A$ is Avogadro's constant, and $M_{\rm Xe}$ is the molar mass of $^{136}$Xe. For the exclusion sensitivity $S_{90\% CL} = 1.64 \sqrt{B}$, while for the 3$\sigma$ discovery potential $S_{3\sigma}$ is constructed using Poisson statistics as detailed in~Ref.~\cite{matteo_fom}. Independently of the applied metric, the background $B$ and the sensitive exposure $\mathscr{E}$ depend on the choice of fiducial volume and energy ROI, both of which can be optimised. 

We find the optimal fiducial volume by gradually adding contiguous detector regions with lowest background index until the sensitivity is maximized. We find optimal $^{nat}$Xe masses of \SI{8.2}{t} (\SI{11.0}{t}) and \SI{13.6}{t} (\SI{17.2}{t}) for the 3$\sigma$-significance discovery potential in the nominal (optimistic) scenario for the 60~t and 80~t configurations, respectively. After this point, $B$ grows faster than $\mathscr{E}$ and the sensitivity decreases gradually. The precise optimum volume depends on the underlying sensitivity metric, with $3\sigma$ discovery potential generally leading to larger fiducial masses compared to the 90\% CL exclusion sensitivity in all studied detector configurations. The external background is dominant in more than 75\% of the active mass, and accordingly the optimised fiducial mass contains less than a quarter of the available target.

The choice of the optimal ROI depends on the spectral shape of the background: on a flat background dominated by the internal background sources, $S / \sqrt{B}$ optimises in a symmetric $\pm 1.4 \sigma$ ROI, yielding a 4\% relative increase in $T_{1/2}^{0\nu}$ sensitivity compared to a $\pm 1\sigma$ ROI; in a background setting dominated by the external $^{214}$Bi $\gamma$-ray peak, an asymmetric ROI is more favourable, with the optimal lower bound depending on the energy resolution. For 0.65\% energy resolution this increase in sensitivity can be as high as 11\% compared to the symmetric $\pm 1 \sigma$. For simplicity and the benefit of a straightforward comparison between different scenarios, we keep the energy ROI fixed at $\pm 1 \sigma$, resulting in a signal detection efficiency of 58\% (which includes the 85\% efficiency from SS/MS discrimination). 

A study of the possible sensitivity gains to be realized with more sophisticated analysis methods such as a profile likelihood ratio (PLR) test, which would allow to exploit a larger fraction of the active mass and a wider energy range, is beyond the scope of this work. In the LZ sensitivity study the use of a PLR-based approach resulted in an improvement of 40\% to $T_{1/2}^{0\nu}$ compared to a counting analysis in an optimised fiducial volume~\cite{lz-xe136-sensitivity}.

\section{\label{sec:discussion}Results and discussion}

\begin{figure}[tbp]
    \begin{indented}
        \item[]\includegraphics[width=0.75\columnwidth]{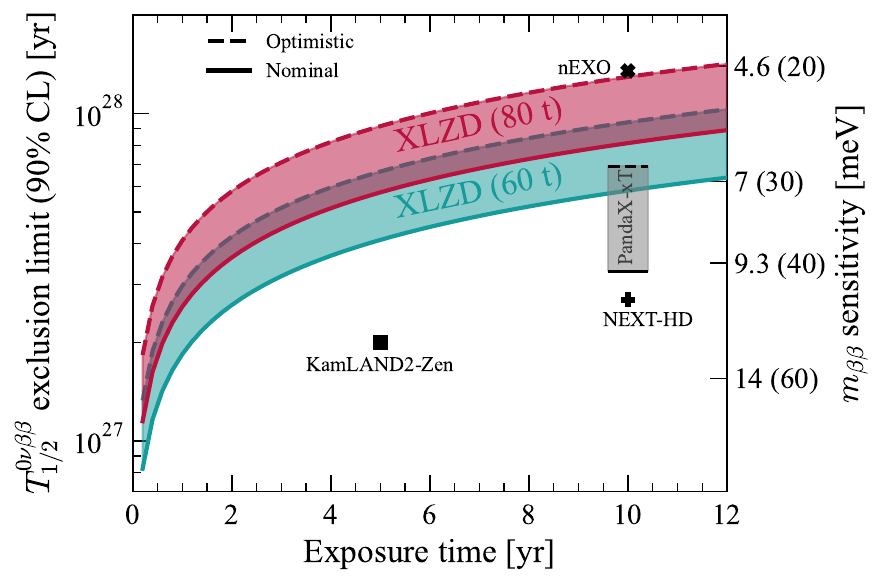}
        \item[]\includegraphics[width=0.75\columnwidth]{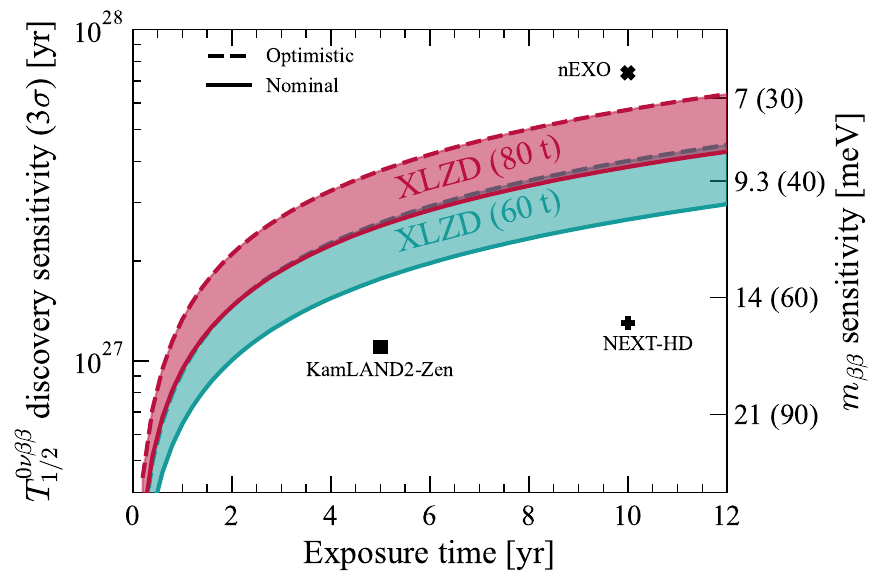}
    \end{indented}
    \caption{Sensitivity of XLZD to the \NDBD~decay of $^{136}$Xe in the two metrics considered in this work: 90\% CL exclusion (top) and $3\sigma$ discovery potential (bottom). The projections for the two mass configurations in the two detector performance scenarios considered are shown by the coloured bands, with the nominal scenario setting the lower band limits and the optimistic scenario the upper limits. Also shown are the projections from other planned experiments: KamLAND2-Zen~\cite{agostini:2023,kamland2-projection}, NEXT-HD~\cite{next-2021}, PandaX-xT~\cite{pandax-xt} and nEXO~\cite{nexo-2022}. Note that the nEXO projections were obtained using a profile likelihood ratio test while those for XLZD use the figure-of-merit estimator. The right axis shows the projected sensitivity to the effective Majorana neutrino mass, $m_{\beta\beta}$, considering a maximum (minimum) $M_{136\rm{Xe}}^{0\nu}$ of 4.77 (1.11) (see text).}
    \label{fig:money_plots}
\end{figure}

The XLZD \NDBD~decay sensitivity projections as a function of exposure time are shown in \cref{fig:money_plots} for both metrics and the two mass configurations being considered. For each of these stages the lower bound of the band corresponds to the nominal scenario and the upper bound to the optimistic scenario. 
In the \SI{80}{t} configuration, and if the optimistic scenario is realised, XLZD can exclude \NDBD~decay half-lives up to $1.3\times10^{28}$~yr at 90\% CL with 10~years of data. The corresponding $3\sigma$ discovery potential is $5.7\times10^{27}$~yr. These projections show that XLZD can surpass currently running and planned experiments using $^{136}$Xe, such as PandaX-xT~\cite{pandax-xt}, KamLAND2-Zen~\cite{kamland2-projection}, NEXT-HD~\cite{next-2021}, and nEXO~\cite{nexo-2022}, while using a target of natural abundance xenon which brings a much broader science program. 

The sensitivity is primarily driven by the target mass, with the external $\gamma$-ray background rate having the largest impact within each mass. For instance, a $\gamma$-ray background at 25\% the level of LZ while maintaining the remaining optimistic scenario assumptions decreases the 3$\sigma$ sensitivity by 15.5\%; installation at LNGS instead of SURF has an 11\% impact; switching the BiPo efficiency, SS/MS discrimination, and energy resolution from the optimistic to the nominal scenario reduces the sensitivity in 6\%, 4.3\%, and 2.5\%, respectively. In fact, a poorer energy resolution of 1\% or a pessimistic BiPo efficiency of 99.9\% both have impacts of only 12\%.

It is foreseen that an interim XLZD configuration with \SI{40}{t} at the same TPC diameter but reduced height will be used for initial technical performance verification and early science~\cite{xlzd-designBook}. Considering a short run of only 3 years for this interim stage and the reduced self-shielding capability resulting from the shallower aspect ratio, XLZD can nevertheless reach a 3$\sigma$ sensitivity of 1.0$\times10^{27}$~yr (2.5$\times10^{27}$~yr 90\% CL exclusion) during this initial run in the optimistic scenario, in line with the DARWIN projections for a detector of similar size~\cite{darwin, darwin-erratum}.

\begin{figure}[tbp]
    \begin{indented}
        \item[]\includegraphics[width=0.77\columnwidth]{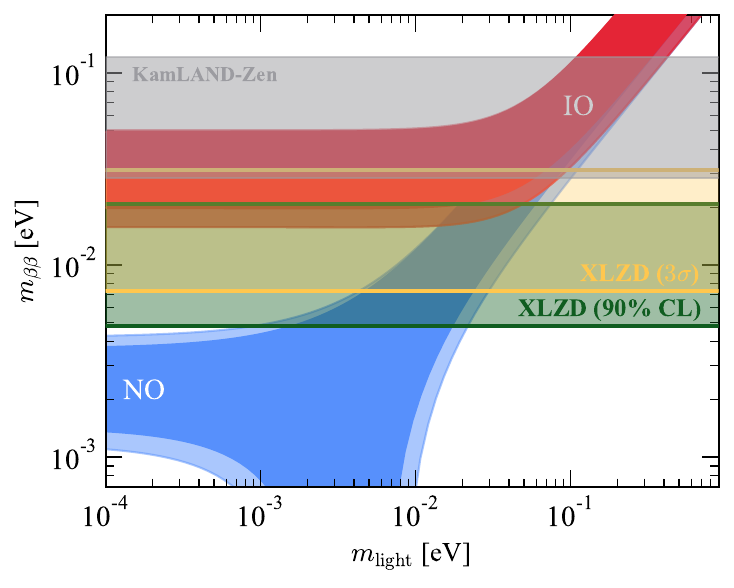}
    \end{indented}
    \caption{Sensitivity of XLZD to the effective Majorana neutrino mass as a function of the lightest neutrino mass for the \SI{80}{t} configuration in the optimistic scenario with 10~years of data. The two metrics considered in this work are shown: $3\sigma$ discovery potential (yellow band) and 90\% CL exclusion (green band). The width of the bands is caused by the uncertainty in the nuclear matrix element models (see main text, also for the expected sensitivity using the most recent nuclear matrix elements). The current best experimental limits from KamLAND-Zen, also from \NDBD~decay in $^{136}$Xe and assuming the same range of NMEs, are shown in grey~\cite{kamland-combined-result}. The allowed regions ($\pm3\sigma$) for the effective Majorana neutrino mass in the inverted (IO) and normal (NO) neutrino mass ordering scenarios are also shown~\cite{nufit-paper,nufit-website}.}
    \label{fig:lobster_plot}
\end{figure}

The right-side axes of the figures in \cref{fig:money_plots} show the XLZD sensitivity to the effective Majorana neutrino mass, $m_{\beta\beta}$, linked to the \NDBD~decay half-life by Eq.~\cref{eq:majorana_mass}. The large uncertainty is dominated by the nuclear matrix element ($M_{136\rm{Xe}}^{0\nu}$). We adopted the parameter values of $g_A = 1.27$, $G^{0\nu}$ from Ref.~\cite{KI_phase_space} and a $M_{136\rm{Xe}}^{0\nu} \in [1.11, 4.77]$ range which covers the predictions from the most commonly used phenomenological models for direct comparison with other experiments~\cite{deformedQRPA,Menendez_2018,Horoi_2016,Coraggio_2020,Mustonen_2013,Hyvarinen_2015,Simkovic_2018,Terasaki_2020,Rodriguez_2010,Song_2017,Barea_2015,Deppisch_2020,nredf}. In particular, the minimum value assumes the deformed-QRPA~\cite{deformedQRPA} model and the maximum the NREDF~\cite{nredf} model. \cref{fig:lobster_plot} shows the allowed $\langle m_{\beta\beta}\rangle$ parameter space as a function of the lightest neutrino mass for the inverted and the normal neutrino mass ordering scenarios, along with the XLZD 90\% CL and $3\sigma$ $\langle m_{\beta\beta}\rangle$ sensitivities for the \SI{80}{t} configuration in the optimistic scenario after 10~years of data, represented by the yellow (4.8--20.5~meV) and green band (7.3--31.3~meV), respectively. XLZD will exclude the inverted neutrino mass ordering---except for the deformed-QRPA model---and probe a significant fraction of the normal ordering scenario. Similarly, if the neutrino masses follow the inverted ordering, XLZD will confirm that hypothesis at the $3\sigma$ level for most of the considered NMEs.

Recent developments in nuclear models include a quenching effect in $g_A$ (required for agreement with experimental data)~\cite{Gysbers2019} and the addition of a previously neglected short-range term~\cite{cirigliano2021-1,cirigliano2021-2}. Considering recent NME models which include these two modifications, the 90\% CL lower limits on $\langle m_{\beta\beta}\rangle$ of XLZD with \SI{10}{years} of data in the \SI{80}{t} configuration and in the optimistic scenario are 7.4--23.5~meV, 4.2--10.2~meV, and 12.1--21.3~meV for the shell model~\cite{Weiss2021,Jokiniemi2022,Castillo2024}, QRPA~\cite{Jokiniemi2022,Castillo2024}, and \textit{ab-initio} calculation~\cite{Belley2023}, respectively. The corresponding 3$\sigma$ discovery potential ranges are 11.2--35.5~meV, 6.3--15.4~meV, and 18.3--32.2~meV, respectively.

\begin{figure}[tbp]
    \begin{indented}
        \item[]\includegraphics[width=0.75\columnwidth]{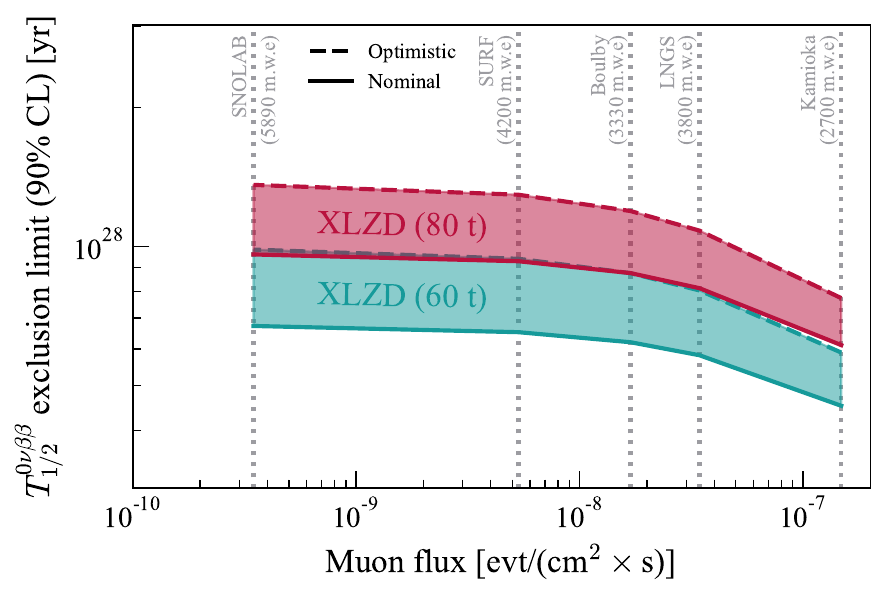}
    \end{indented}
    \caption{Projected 90\% confidence level (CL) exclusion limits of XLZD to the \NDBD~decay of $^{136}$Xe as a function of muon flux (which drives the $^{137}$Xe background) after 10 years of data. The coloured bands show the projections for the two mass configurations, bounded by the nominal (lower limit) and optimistic (upper limit) scenarios (see \cref{table:scenarios}). The vertical dashed lines indicate the fluxes corresponding to each of the possible hosting laboratories, along with their corresponding water equivalent depth.}
    \label{fig:lab_dependence}
\end{figure}

We also studied the effect of the $^{137}$Xe production rate on the sensitivity, as this background will depend on the choice of underground laboratory. The results for the exclusion sensitivity are shown in \cref{fig:lab_dependence}. The bands for each mass stage represent the range between the detector performance scenarios, dominated by the external $\gamma$-ray background. Installation at SNOLAB, with a significantly lower muon rate compared to SURF (optimistic scenario), has only a modest effect in the sensitivity (which increases to 1.4$\times$10$^{28}$~yr for the \SI{80}{t} mass configuration). It is clear that when the cosmogenic $^{137}$Xe rate is sub-dominant (SNOLAB, SURF, Boulby) or at the same level (LNGS) as the $^8$B background, the impact on the sensitivity is much smaller than that of varying the external $\gamma$-ray rate between the two scenarios considered. This is not the case for the Kamioka site, in which case the effect on the projected sensitivity is significant.

\begin{figure}[tbp]
    \begin{indented}
        \item[]\includegraphics[width=0.75\columnwidth]{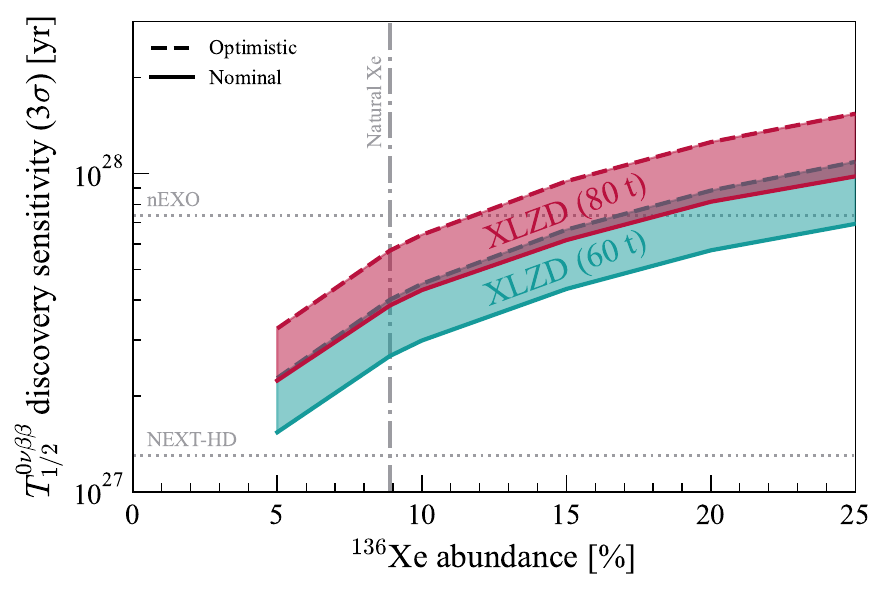}
    \end{indented}
    \caption{Dependence of the \NDBD~3$\sigma$ discovery potential of XLZD with the abundance of $^{136}$Xe in the target with 10~years of data. The coloured bands show the projections for the two mass configurations, and are limited by the nominal (lower limit) and optimistic (upper limit) scenarios (see \cref{table:scenarios}). Also shown are the projections from NEXT-HD~\cite{next-2021} and nEXO\cite{nexo-2022}.}
    \label{fig:enr_dependence}
\end{figure}

XLZD will nominally use natural abundance xenon for its target, but scenarios with different $^{136}$Xe concentration may be warranted at a later stage to investigate a putative signal (in XLZD or in another experiment). \cref{fig:enr_dependence} shows the effect of varying the $^{136}$Xe concentration on the 3$\sigma$ discovery potential for the two mass stages and two detector performance scenarios considered. $^{137}$Xe production was estimated for LNGS~\cite{cuenca-private} (nominal performance) for the various enrichment scenarios and scaled for SURF (optimistic) using the production ratio for natural xenon in these two laboratories (\cref{table:cosmogenics}). If XLZD or nEXO were to report an observation which is in tension with their background model but falling short of a discovery, XLZD will be able to confirm it at the 3$\sigma$ level in 10~years with less than 20\% enrichment in the 80~t stage, or even in the 60~t stage with 25\% enrichment. Conversely, a signal observation in XLZD can also be tested by running the experiment with some depletion level, which would not impact the background expectation significantly while reducing the signal rate---a strategy already being explored by NEXT~\cite{next_bg_subtraction}.

\section{\label{sec:conclusions}Conclusions}

The XLZD collaboration is designing an experiment based on two-phase xenon TPC technology capable of completely probing the spin-independent WIMP-nucleon scattering parameter space down to the neutrino fog. With a target mass of up to 80~t and an extremely low background this experiment will be able to study other well-motivated physics channels. In this work we presented the initial sensitivity projections of XLZD in the search for \NDBD~decay in $^{136}$Xe, a process which can be used to search for new physics, probe the Majorana nature of the neutrino, and determine the neutrino mass ordering.

The large target volume is very effective at shielding the inner region of the detector from external high-energy $\gamma$-rays that would otherwise dominate the background in this search, while providing several tons of the source isotope even at natural abundance. Moreover, the excellent energy resolution ($\sigma_E$ = 0.67\%) already demonstrated in detectors based on dual-phase xenon TPCs can be used to minimize the width of the ROI, thus decreasing leakage of the $^{136}$Xe \DBD~continuum and from the nearby $\gamma$-ray lines from $^{214}$Bi and $^{208}$Tl. Importantly, the $\gamma$-ray background can be further reduced by the capability to identify multiple scatter interactions down to \SI{3}{mm} separation in the vertical direction and with a very low energy threshold per vertex. Finally, an instrumented xenon skin and an outer detector surrounding the TPC provide an additional reduction of this background by vetoing coincident signals, which is particularly effective in mitigating against the 2615~keV line from $^{208}$Tl.

The use of a natural abundance xenon target has the advantage of limiting the cosmogenic production of the $^{137}$Xe background, as $^{136}$Xe has the lowest neutron capture cross section amongst all naturally-occurring xenon isotopes. This background remains sub-dominant if the experiment is installed at SNOLAB, SURF, Boulby or LNGS.

With a thorough material screening and selection campaign and an online radon reduction system capable of reducing $^{222}$Rn to the \SI{0.1}{\micro Bq/kg} level, XLZD will reach a 90\% CL half-life exclusion sensitivity of 1.3$\times$10$^{28}$~yr with 10 years of data in the \SI{80}{t} configuration, fully excluding the inverted neutrino mass ordering scenario for all but one of the most commonly used NMEs. With a 3$\sigma$ discovery sensitivity of 5.7$\times$10$^{27}$~yr it will also probe the inverted ordering scenario for a signal for most of these NME models. 
These sensitivities will be further improved by the use of a PLR-based statistical analysis in a larger volume and using a wider energy range, realistically allowing the full exclusion of the inverted ordering at 90\% CL. Conversely, recent NME calculations suggest somewhat reduced sensitivities for all $0\nu\beta\beta$~decay experiments~\cite{agostini:2023,Weiss2021,Jokiniemi2022,Belley2023,Castillo2024}.
A second stage can be envisioned using xenon with some level of $^{136}$Xe enrichment to increase its sensitivity, allowing for probing at the 3$\sigma$ level a possible hint of a signal during the first XLZD stage at the $T_{1/2} = 1.3\times10^{28}$~yr level.

Conversion from measured half-lives to the physically relevant effective Majorana neutrino mass requires knowledge of the NME for the relevant isotope, which can vary by factors of a few between nuclear models resulting in large uncertainties in $\langle m_{\beta\beta}\rangle$. XLZD will be able to measure or significantly improve the current best limits on the half-life of the $^{136}$Xe \DBD~decay to the first excited 0$^+$ state of $^{136}$Ba~\cite{exo200_excited,kamland-excited}. This yet unobserved SM-allowed decay can thus be used to benchmark the predictions of the various nuclear models and help to reduce theoretical uncertainties~\cite{2vbb_excited}.
Complementarity between experiments using different isotopes (e.g.~SNO+~\cite{sno+}, LEGEND~\cite{legend}, CUPID~\cite{cupid}, AMoRE~\cite{amore}, SuperNEMO\cite{supernemo}) is crucial to reduce NME uncertainties when probing the neutrino mass hierarchy space and to claim a possible discovery. Furthermore, multiple measurements in different isotopes will be required to characterise the physical mechanism mediating this decay.

\vspace{0.5cm}
\section*{Acknowledgements}
This work was supported by the UKRI's Science \& Technology Facilities Council under awards ST/W000636/1, ST/Y003594/1, ST/Z000866/1, ST/W000547/1, ST/Y003586/1, ST/Z000807/1, ST/Y003527/1, ST/Z00084X/1, ST/W00058X/1, and ST/Z000831/1; the U.S. Department of Energy under contract numbers DE-SC0014223, DE-SC0012447, DE-AC02-76SF00515, and DE-SC0012704; the U.S. National Science Foundation under award number 2112802; the German Federal Ministry of Education and Research (BMBF) under the ErUM-Pro framework grant no.~05A23UM1, 05A23VF1 (CRESST-XENON-DARWIN), 05A23PM1, and 05A23VK3; the German Research Foundation (DFG) through the Research Training Group 2149 and the PRISMA\textsuperscript{+} Excellence Cluster project no.~390831469; the program ``Matter and the Universe" of the Helmholtz Association, through the Helmholtz Initiative and Networking Fund (grant no.~W2/W3-118); the European Research Council (ERC) by the ERC Advanced Grants Xenoscope (no.~742789) and LowRad (no.~101055063), as well as by the European Union’s Horizon 2020 research and innovation programme under grant agreements 724320 (ULTIMATE), 101020842, and Marie Skłodowska-Curie grant agreement No 860881-HIDDeN; the MCIN/AEI/10.13039/501100011033 from the following grants: PID2020-118758GB-I00, CNS2022-135716 funded by the ``European Union NextGenerationEU/PRTR" and CEX2019-000918-M to the ``Unit of Excellence Mar\'ia de Maeztu 2020-2023" award to the Institute of Cosmos Sciences; the Swiss National Science Foundation under grant numbers PCEFP2-181117, 20FL20-216573, 200020-219290; the Next Generation EU, Italian National Recovery and Resilience Plan, Investment PE1–``Future Artificial Intelligence Research''; the Fondazione ICSC, Spoke 3 ``Astrophysics and Cosmos Observations'', the Italian National Recovery and Resilience Plan (Project ID CN00000013); the Italian Research Center on High-Performance Computing, Big Data and Quantum Computing funded by MUR Missione 4 Componente 2 Investimento 1.4.; the Istituto Nazionale di Fisica Nucleare (Italy); the Portuguese Foundation for Science and Technology (FCT) under award number PTDC/FIS-PAR/2831/2020; the JSPS Kakenhi under contract numbers 24H02240, 24H02236, 24K00659, and 24H00223 and by JST FOREST Program under contract number JPMJFR212Q; the Dutch Research Council (NWO) and the Dutch national e-infrastructure with the support of SURF Cooperative; the Israeli Science Foundation 1859/22 and the Weizmann Institute; the Australian Government through the Australian Research Council Centre of Excellence for Dark Matter Particle Physics (CDM, CE200100008); the Region des Pays de la Loire and the Australia-France Network of Doctoral Excellence (AUFRANDE) program. 
For the purpose of open access, the authors have applied a Creative Commons Attribution (CC BY) licence to any Author Accepted Manuscript version arising from this submission.

\section*{References}
\bibliography{bibliography}

\end{document}